\documentclass[12pt,a4paper]{iopart}

\usepackage[english]{babel}
\usepackage [latin1]{inputenc}

\usepackage{iopams}
\usepackage{verbatim}

\usepackage{graphicx}
\usepackage{float}
\usepackage{placeins}

\usepackage{color}

\usepackage[colorlinks,citecolor=blue,linkcolor=black,urlcolor=blue]{hyperref}

\usepackage{subcaption} 

\def\iotabar{\lower3pt\hbox{$\mathchar'26$}\mkern-7mu\iota}
\newcommand {\aplt} {\ {\raise-.5ex\hbox{$\buildrel<\over\sim$}}\ }
\newcommand{\dd}{\mbox{d}}

\newcommand{\eq}[1]{(\ref{#1})}
\newcommand{\bun}{\hat{\mathbf{b}}}

\newcommand{\bv}{\mathbf{v}}
\newcommand{\bB}{\mathbf{B}}

\newcommand{\bk}{\mathbf{k}}

\newcommand{\dotcross}{ \raise 0.65ex\hbox{${\scriptstyle {{_{\displaystyle \cdot}}\atop\times}}$} }
\newcommand{\crossdot}{ \raise 0.5ex\hbox{${\scriptstyle {{_\times}\atop{\displaystyle \cdot}}}$} }

\newcommand{\kappabf}{\mbox{\boldmath$\kappa$}}

\newcommand{\sumsig}{ \raise -1.3ex\hbox{${{\displaystyle \sum}\atop{\scriptstyle \sigma}}$} }
\newcounter{appnumb}

\newcommand{\nablam}{\boldsymbol{\nabla}\hspace*{-0.2mm}}
\newcommand{\nablammm}{\boldsymbol{\nabla}\hspace*{-0.4mm}}

\begin{document}

\title[Modification of ITG turbulence by impurities in stellarator plasmas] {Modification of ion-temperature-gradient turbulence by impurities in stellarator plasmas}


\author{Iv\'an Calvo$^{1}$, F\'elix I. Parra$^{2}$, Hanne Thienpondt$^{1}$ and Jos\' e Manuel Garc\'{\i}a-Rega\~na$^{1}$
}

\vspace{0.2cm}

\address{$^1$Laboratorio Nacional de Fusi\'on, CIEMAT, 28040 Madrid, Spain}
\address{$^2$Princeton Plasma Physics Laboratory, Princeton, NJ 08543, USA}




\vskip 0.5cm

{\large
\begin{center}
\today
\end{center}
}

\vspace{0.0 cm}

\begin{abstract}
Recent nonlinear gyrokinetic simulations have shown that impurities can strongly modify the turbulent heat flux in stellarator plasmas. Here, the ion-temperature-gradient (ITG) dispersion relation in a plasma containing impurities is analytically solved in certain limits and an expression for the modification of the ITG growth rate by impurities is derived. The analytical expression is the sum of three terms corresponding to three different physical causes (impurity density gradient, impurity temperature gradient and dilution) of the change in the growth rate. The scalings predicted analytically for the modification of the growth rate are shown to be reproduced by linear gyrokinetic simulations. The conditions for reduction or increase of the ITG growth by impurities are also correctly predicted by the analytical solution to the dispersion relation. Finally, a remarkable correlation is found between the analytical expression for the modification of the growth rate and the modification of the turbulent heat flux obtained from nonlinear gyrokinetic simulations.
\end{abstract}


\section{Introduction}
\label{sec:introduction}

In general, the combination of small collisionality and the three-dimensionality of stellarator magnetic fields produces very large neoclassical transport~\cite{Helander2012, Calvo_2017, Herbemont_2022}. This is why neoclassical mechanisms have typically dominated energy and particle transport in the core of stellarator plasmas~\cite{Dinklage_nf_2013}. This has changed with the arrival of neoclassically optimized stellarators: once neoclassical transport is reduced, turbulent transport plays a more prominent role. In Wendelstein 7-X (W7-X)~\cite{Bosch2017, Wolf_nf_2017, Klinger2019}, the first large stellarator designed to have small neoclassical transport, ion-temperature-gradient (ITG) driven turbulence is often responsible for most of the energy transport throughout the plasma volume~\cite{Bozhenkov2020} and limits the ion temperature in the core~\cite{Beurskens_2021}.

In standard electron-cyclotron-resonance-heated plasmas, scenarios with reduced plasma turbulence have been transiently reached by injection of cryogenic pellets. The neoclassical optimization of W7-X has been experimentally demonstrated in these scenarios~\cite{Beidler2021}. The study of turbulence reduction mechanisms leading to enhanced-performance scenarios is expected to be an active research area in the near future.

Here, we concentrate on impurities as a different mechanism for the modification of turbulent plasma transport and, in particular, for its reduction. Experimentally, impurities have been observed to reduce turbulence in tokamaks~\cite{McKee2000} and stellarators~\cite{Osakabe2014, Takahashi2018, Lunsford2021, Nespoli2022}. In experiments it is difficult to assess the importance of the different factors that can lead to turbulence 	reduction, and hence it is difficult to extrapolate their relevance for reactors. In the realm of numerical simulations, and with a focus on stellarators, the question of how impurities modify the turbulent heat fluxes of the bulk species  when the profiles of the bulk ions are kept constant has recently been addressed in~\cite{GarciaRegana2024}, showing that impurities can strongly modify the bulk-ion heat flux. In particular, it is shown that impurities can increase or reduce the heat flux depending, essentially, on the relative sign of the bulk-ion and impurity density gradients. In the present paper, we aim to provide theoretical insight into some of the main results of \cite{GarciaRegana2024}.  Whereas the gyrokinetic simulations in \cite{GarciaRegana2024} are nonlinear, electrostatic and include kinetic electrons, bulk ions and impurities, we will carry out our theoretical discussion for ITG turbulence with adiabatic electrons. In the literature, the influence of impurities on ITG turbulence has been addressed in simpler magnetic geometries such as a sheared-slab or a tokamak, and mostly by means of linear simulations~\cite{Tang1980, Dominguez1993, Paccagnella1990, Frojdh1992, Dong1994, Dong1995, Du2014, Du2015} or simple nonlinear models~\cite{Fu1997}. Our approach encompasses linear and nonlinear gyrokinetic simulations in stellarator geometry, as well as analytical solutions to the ITG dispersion relation in certain tractable limits. We will see that the main effects reported in \cite{GarciaRegana2024} are well captured even by linear analytical calculations based on approximations to the gyrokinetic equations, which is an additional argument for the robustness of such effects.

The rest of the paper is organized as follows. In section \ref{sec:gk_eqs}, we give the gyrokinetic equations for ITG turbulence in stellarator plasmas including impurities, and their linearization. In section \ref{sec:magnetic geometries}, we present the magnetic geometries employed to illustrate numerical and analytical results throughout the paper, corresponding to W7-X and LHD configurations. In section \ref{sec:local_models}, the so-called toroidal ITG and other local models are introduced. We argue that in relevant regions of parameter space, ITG modes are localized along magnetic field lines, and local models give reasonable approximations to the growth rate and frequency of the modes. The discussion on mode localization provides some justification for the analytical calculation of section \ref{sec:solution_ITG_dispersion_relation}, where the toroidal ITG dispersion relation is solved in certain asymptotic limits. Specifically, we give an explicit expression for the modification of the growth rate by impurities. The formula for the modification of the toroidal ITG growth rate is the sum of three terms corresponding to the effect of the impurity density gradient, the effect of the impurity temperature gradient and the effect of dilution (i.e.~of adding impurities with vanishing density and temperature gradients). The analytical calculation gives the conditions for which impurities reduce or increase the ITG growth rate. In section \ref{sec:analytical_expression_vs_linear_simulations}, we show that the scalings predicted by the analytical expression for the modification of the ITG growth rate agree with those obtained from linear gyrokinetic simulations for realistic values of the parameters and, in particular, of the effective charge. In section \ref{sec:analytical_expression_vs_nonlinear_simulations}, it is shown that there exists a remarkable correlation between the analytical expression for the modification of the growth rate and the modification of the turbulent ion heat flux obtained from the nonlinear gyrokinetic simulations in reference \cite{GarciaRegana2024}. The conclusions are given in section \ref{sec:conclusions}.

\section{Gyrokinetic equations for electrostatic turbulence in stellarators}
\label{sec:gk_eqs}

Let us introduce the equations that describe electrostatic turbulence in strongly magnetized plasmas confined in a stellarator magnetic field with nested flux surfaces\footnote{Along the paper, we will use the terms \emph{flux surfaces} and \emph{magnetic surfaces} interchangeably.}. Strong magnetization implies $\rho_{s*} := \rho_s/a \ll 1$, where $\rho_s = v_\perp/\Omega_s$ is the Larmor radius (also called gyroradius), $v_\perp$ is the component of the velocity perpendicular to the magnetic field $\bB$, $\Omega_s$ is the gyrofrequency, $a$ is the stellarator minor radius and $s$ is a species index that can take the values $s = e, i, z$ for electrons, bulk ions and impurity ions, respectively. For simplicity, we assume that there is a single impurity species, although all the results of the paper can easily be generalized if there are multiple impurity species. The description of turbulence at the low collisionalities typical of fusion plasmas requires a kinetic treatment, but thanks to the smallness of the normalized gyroradius, one can derive reduced kinetic equations by averaging out, order by order in an expansion in $\rho_{s*}  \ll 1$, the fast gyration of the charged particles around magnetic field lines. Gyrokinetics~\cite{Catto1978, Frieman1982} is the theory that gives the procedure to obtain these reduced equations, known as gyrokinetic equations, that manifestly exhibit the strong anisotropy of the turbulence along and across the magnetic field that is characteristic of these plasmas. In subsection \ref{sec:fluxtube_gk_eqs}, we introduce the collisionless electrostatic gyrokinetic equations in the flux-tube approximation~\cite{Beer1995}. In subsection \ref{sec:lin_gk_eqs}, we give the linearization of these equations, that will be used extensively in the paper. In subsection \ref{sec:gk_eq_simplified_for_ITG}, we point out how the equations of subsections \ref{sec:fluxtube_gk_eqs} and \ref{sec:lin_gk_eqs} are particularized for ITG turbulence.

\subsection{Flux-tube gyrokinetic equations for electrostatic turbulence}
\label{sec:fluxtube_gk_eqs}

First, we specify our choice of phase-space coordinates. As spatial coordinates we employ $\{r,\alpha,\ell\}$, where $r\in[0,a]$ is a radial coordinate labeling magnetic surfaces, $\alpha\in[0,2\pi)$ is a poloidal angle that labels field lines on each magnetic surface, $\ell \in[0,\ell_{\rm max}(r,\alpha)]$ is the arc length along magnetic field lines and $\ell_{\rm max}(r,\alpha)$ is the length of the field line after completing a toroidal turn. In these coordinates, the magnetic field reads
\begin{equation}\label{eq:def_magnetic_field}
\bB = \Psi_t'\nabla r \times \nabla\alpha,
\end{equation}
with $2\pi\Psi_t(r)$ the toroidal flux and primes standing for derivatives with respect to $r$. The volume element $\sqrt{g} = [(\nabla r\times\nabla\alpha)\cdot\nabla \ell]^{-1}$ has the form
\begin{equation}\label{eq:sqrt_g}
\sqrt{g} = \frac{\Psi'_t}{B},
\end{equation}
with $B = |\bB|$ the magnetic-field strength. Note that \eq{eq:def_magnetic_field} and \eq{eq:sqrt_g} imply
$\bun\cdot\nabla \ell = 1$, where $\bun = B^{-1} \bB$. As velocity coordinates we use $u$ and $\mu$, where $u = \bun\cdot\bv$ is the component of the velocity $\bv$ parallel to $\bB$ and $\mu = v_\perp^2/2B$ is the magnetic moment per mass unit.
 
The distribution function of species $s$ is denoted by $F_s$ and expanded in $\rho_{s*} \ll 1$ as
\begin{equation}
F_s = f_{Ms} + f_{s1} + \dots,
\end{equation}
where $f_{s1} (r,\alpha,\ell,u,\mu,t) \sim O(\rho_{s*} f_{Ms})$ is the turbulent perturbation to the Maxwellian distribution
\begin{equation}
\fl
f_{Ms}(r,\alpha,\ell,u,\mu) = n_s(r)\left(\frac{m_s}{2\pi T_s(r)}\right)^{3/2}\exp\left( - \frac{m_s(u^2/2 +\mu B(r,\alpha,\ell))}{T_s(r)}\right).
\end{equation}
Note that $f_{s1}$ depends on the time $t$, but $f_{Ms}$ does not. The lowest-order density $n_s$ and temperature $T_s$ are constant on flux surfaces, and quasineutrality implies, to lowest order,
\begin{equation}
\sum_s Z_s e n_s = 0,
\end{equation}
where $e$ is the proton charge and $Z_s e$ is the charge of species $s$. In general, collisional coupling between bulk and impurity ions leads to $T_z = T_i$, although we will not assume this in our derivations.

The fields to be determined from the gyrokinetic equations are $f_{s1}$ and the turbulent electrostatic potential, that we denote by $\varphi(r, \alpha, \ell, t)$. In the gyrokinetic ordering, $f_{s1}$ and $\varphi$ vary on small scales in $r$ and $\alpha$, and on large scales in $\ell$. In order to fully exploit this scale separation, $f_{s1}$ and $\varphi$ are Fourier expanded in the variation on small scales,
\begin{eqnarray}\label{eq:Fourier_expansions_f_and_varphi}
f_{s1} = \sum_{k_r, k_\alpha} \hat f_{s1} (r,\alpha,\ell,k_r,k_\alpha,u,\mu,t)
\exp\left(
i k_r r + i k_\alpha \alpha
\right)
,
\nonumber\\[5pt]
\varphi = \sum_{k_r, k_\alpha} \hat \varphi (r,\alpha,\ell,k_r,k_\alpha,t)
\exp\left(
i k_r r + i k_\alpha \alpha
\right).
\end{eqnarray}
The orderings just mentioned are formalized by assuming that $\hat f_{s1} $ and $\hat \varphi$ vary on spatial scales $O(a)$ and that the summation in \eq{eq:Fourier_expansions_f_and_varphi} is over wavenumbers $O(\rho_s^{-1})$.

The set of flux-tube gyrokinetic equations consists of the gyrokinetic Vlasov equation,
\begin{eqnarray}\label{eq:h_equation_local}
\fl
\partial_t \left(
\hat h_s - \frac{Z_s e}{T_s} \hat \varphi J_0(k_\perp \rho_s) f_{Ms}
\right) + (u \partial_\ell - \mu \partial_\ell B\partial_u) \hat h_s
+ i \bk_\perp\cdot \bv_{Ms} \hat h_s
\nonumber\\[5pt]
\fl
\hspace{1cm}
-\frac{1}{B}\sum_{k'_r, k'_\alpha, k''_r, k''_\alpha}
(\bk'_\perp \times \bk''_\perp)\cdot\bun \,
\hat \varphi (k'_r,k'_\alpha) J_0(k'_\perp \rho_s)
\hat h_s(k''_r,k''_\alpha)
\Bigg\vert_{\bk'_\perp + \bk''_\perp = \bk_\perp}
\nonumber\\[5pt]
\fl
\hspace{1cm}
- \frac{ik_\alpha}{\Psi'_t} \hat\varphi J_0(k_\perp \rho_s)
\left[
\frac{n'_s}{n_s} + \frac{T'_s}{T_s}
\left(
\frac{m_s(u^2/2 + \mu B)}{T_s} - \frac{3}{2}
\right)
\right] f_{Ms}
 = 0,
\end{eqnarray}
and the gyrokinetic quasineutrality equation,
\begin{eqnarray}\label{eq:quasineutrality_h}
\fl
\sum_s 2\pi Z_s B\int_{-\infty}^\infty\dd u\int_0^\infty\dd\mu \,
\hat h_s 
J_0(k_\perp\rho_s)
- \sum_s \frac{Z_s^2 e n_s}{T_s}
\hat \varphi
= 0,
\end{eqnarray}
where
\begin{equation}
\hat h_s := \hat f_{s1} + \frac{Z_s e}{T_s} \hat \varphi J_0(k_\perp \rho_s) f_{Ms}
\end{equation}
is the non-adiabatic component of $\hat f_{s1}$, $J_0(\cdot)$ is the zeroth order Bessel function of the first kind,
\begin{equation}
\bv_{Ms} = \frac{1}{\Omega_s}\bun\times\left(u^2 \kappabf + \mu \nabla B\right)
\end{equation}
is the magnetic drift, $\kappabf = \bun\cdot\nabla\bun$ is the magnetic curvature, $\Omega_s = Z_s e B / m_s$,
\begin{equation}
\bk_\perp := k_r\nabla r + k_\alpha \nabla\alpha
\end{equation}
 is the perpendicular wave vector and $k_\perp := |\bk_\perp|$. In equations \eq{eq:h_equation_local} and \eq{eq:quasineutrality_h}, and in the rest of the paper, we often ease the notation by displaying only some of the variables on which functions such as $\hat h_s$ and $\hat \varphi$ depend.
  
\subsection{Linearization of the gyrokinetic equations}
\label{sec:lin_gk_eqs}

If we drop the nonlinear term of \eq{eq:h_equation_local}, we obtain
\begin{eqnarray}\label{eq:h_equation_local_linear_stella}
\fl
\partial_t \left(
\hat h_s - \frac{Z_s e}{T_s} \hat \varphi J_0(k_\perp \rho_s) f_{Ms}
\right) + (u \partial_\ell - \mu \partial_\ell B\partial_u) \hat h_s
+ i \bk_\perp\cdot \bv_{Ms} \hat h_s
\nonumber\\[5pt]
\fl
\hspace{1cm}
- \frac{ik_\alpha}{\Psi'_t} \hat\varphi J_0(k_\perp \rho_s)
\left[
\frac{n'_s}{n_s} + \frac{T'_s}{T_s}
\left(
\frac{m_s(u^2/2 + \mu B)}{T_s} - \frac{3}{2}
\right)
\right] f_{Ms}
 = 0.
\end{eqnarray}
Equations \eq{eq:h_equation_local_linear_stella} and \eq{eq:quasineutrality_h} are the linearization of the set of equations \eq{eq:h_equation_local} and \eq{eq:quasineutrality_h}.

It is often useful to Fourier transform $\hat h_s$ and $\hat \varphi$ in time and solve the linearized set of equations for each component of the Fourier decomposition. Assuming
\begin{eqnarray}
&&\hat h_s(r,\alpha,\ell,k_r,k_\alpha,u,\mu,t) = e^{-i \omega t} \check h_s(r,\alpha,\ell,k_r,k_\alpha,u,\mu,\omega),
\nonumber \\[5pt]
&&\hat \varphi (r,\alpha,\ell,k_r,k_\alpha,t) = e^{-i \omega t} \check \varphi (r,\alpha,\ell,k_r,k_\alpha,\omega),
\end{eqnarray}
and using these expressions in \eq{eq:h_equation_local_linear_stella} and \eq{eq:quasineutrality_h}, one obtains
\begin{equation}\label{eq:h_equation_local_linear}
\fl
i (u \partial_\ell - \mu \partial_\ell B\partial_u) \check h_s
+
\left(\omega - \omega_{ds} \right)
\check h_s
=
\left(\omega -  \omega_{*s}^T  \right)
\frac{Z_s e}{T_s} \check \varphi J_0(k_\perp \rho_s) f_{Ms}
\end{equation}
and
\begin{equation}\label{eq:quasineutrality_h_linear}
\fl
\sum_s 2\pi Z_s B\int_{-\infty}^\infty\dd u\int_0^\infty\dd\mu \,
\check h_s 
J_0(k_\perp\rho_s)
-
\sum_s \frac{Z_s^2 e n_s}{T_s}
\check \varphi
=
0
.
\end{equation}
Solving \eq{eq:h_equation_local_linear} and \eq{eq:quasineutrality_h_linear} for all values of $\omega$ is equivalent to solving \eq{eq:h_equation_local_linear_stella} and \eq{eq:quasineutrality_h} for each value of $t$. In \eq{eq:h_equation_local_linear} we have introduced some standard notation. Specifically,
\begin{equation}
\omega = {\rm Re}(\omega) + i \gamma
\end{equation}
is the complex frequency of the linear mode, ${\rm Re}(\omega)$ is the real frequency, $\gamma = {\rm Im}(\omega)$ is the growth rate,
\begin{equation}
\omega_{ds} :=  \bk_\perp\cdot \bv_{Ms}
\end{equation}
is the drift frequency,
\begin{equation}
\eta_s :=  \frac{(\ln T_s)'}{(\ln n_s)'},
\end{equation}
\begin{equation}\label{eq:def_omegastarsT}
\omega_{*s}^T :=  \omega_{*s} \left[
1 + \eta_s
\left(
\frac{m_s(u^2/2 + \mu B)}{T_s} - \frac{3}{2}
\right)
\right]
\end{equation}
and
\begin{equation}
\omega_{*s} :=  \frac{T_s k_\alpha}{Z_s e \Psi'_t} \frac{n'_s}{n_s}
\end{equation}
is the diamagnetic frequency. As is customary, given a profile $X(r)$, we define its variation scale length $L_X$ at the radial position $r$ by the relation $a/L_X = - X'/X$.

Note that if a gyrokinetic code that solves an initial value problem is employed to solve the set of equations \eq{eq:h_equation_local_linear_stella} and \eq{eq:quasineutrality_h}, its solution will tend, as $t\to\infty$, to the solution of the set of equations \eq{eq:h_equation_local_linear} and \eq{eq:quasineutrality_h_linear} with the largest value of $\gamma$; that is, to the so-called fastest-growing mode.

\subsection{Particularization of the equations of subsections \ref{sec:fluxtube_gk_eqs} and \ref{sec:lin_gk_eqs} for ITG turbulence}
\label{sec:gk_eq_simplified_for_ITG}

The theoretical discussion and the analytical calculations below will be carried out in the simpler framework of ITG turbulence. The ITG equations are obtained by assuming in \eq{eq:h_equation_local} and \eq{eq:quasineutrality_h} that the electron response is adiabatic,
\begin{equation}
\hat h_e \equiv 0.
\end{equation}
Hence, apart from the gyrokinetic quasineutrality equation \eq{eq:quasineutrality_h} with $\hat h_e \equiv 0$, we are left with two gyrokinetic Vlasov equations of the form  \eq{eq:h_equation_local}, one for $\hat h_i$ and one for $\hat h_z$. The same applies to the linearization of the equations.

\section{Magnetic geometries employed in the paper}
\label{sec:magnetic geometries}

\begin{figure}[h!]
	\begin{subfigure}{0.5\linewidth}
		\includegraphics[width=0.98\textwidth] {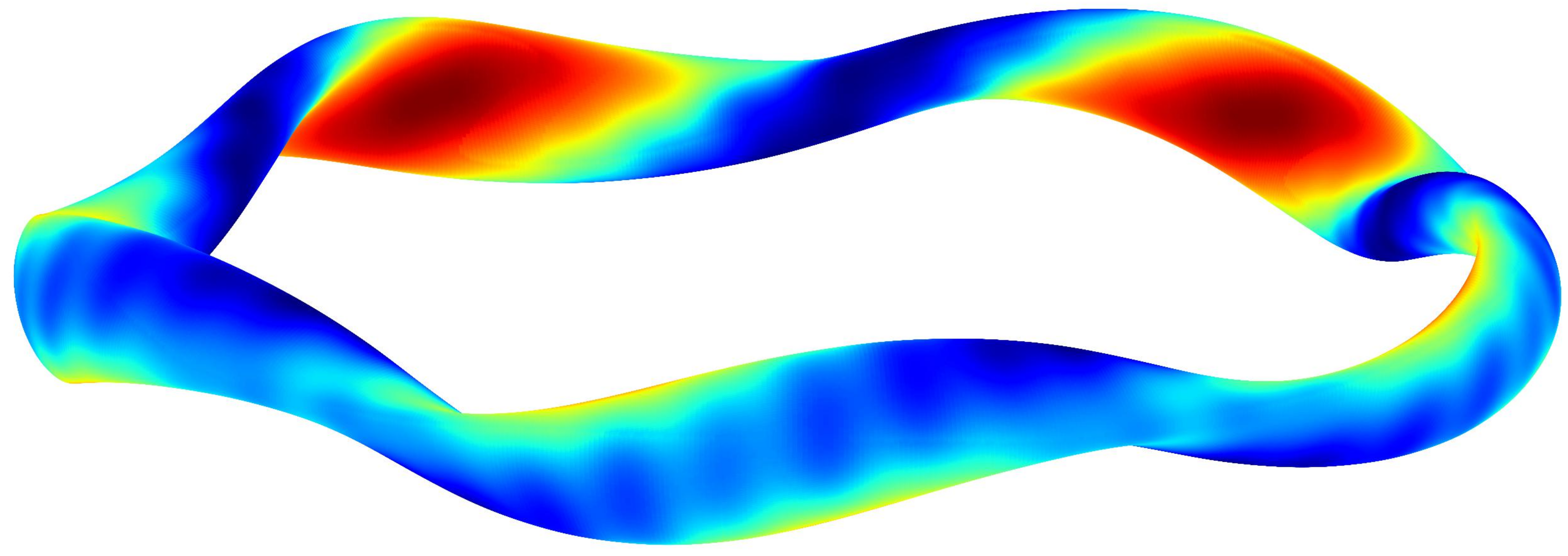}
	\end{subfigure}%
	\begin{subfigure}{0.5\linewidth}
		\hspace*{-1mm}
		\vspace*{-5mm}
		\includegraphics[trim={110mm 0mm 110mm 0mm}, clip, width=0.98\textwidth] {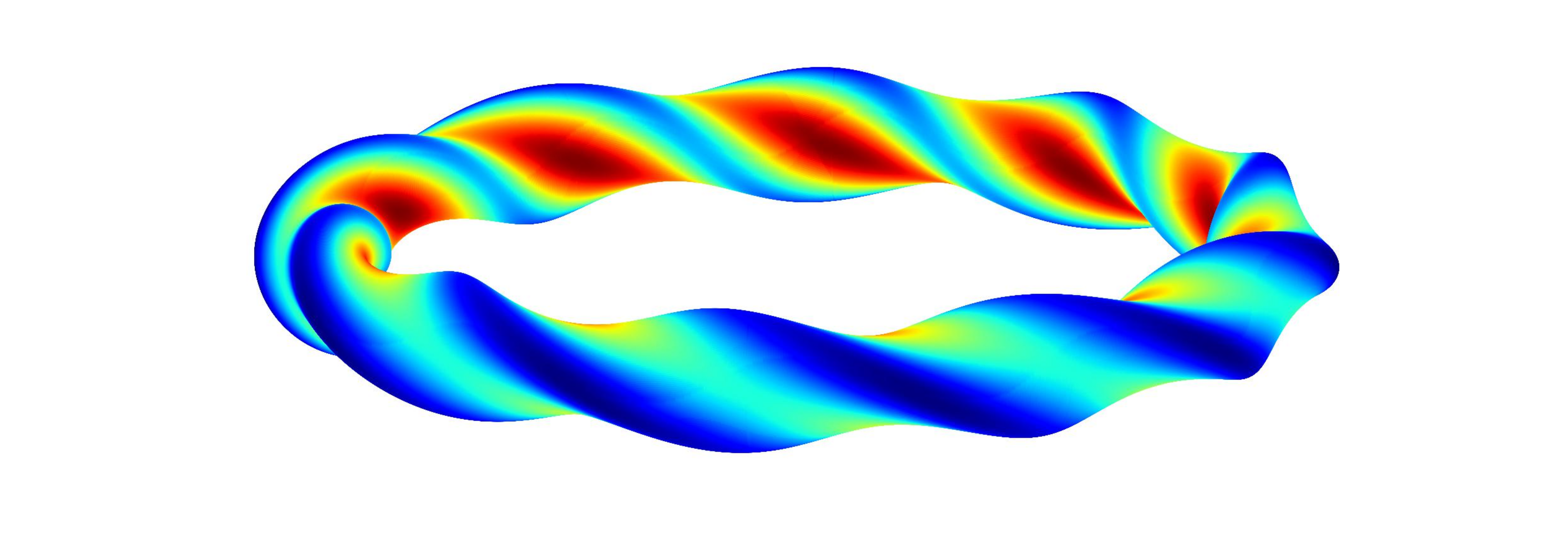}
	\end{subfigure}
\centering
\caption{Flux surfaces $r/a = 0.7$ of the standard configuration of W7-X (left) and an inward-shifted configuration of LHD (right). The color represents the value of $B$, with red corresponding to the largest values and blue to the smallest values.}
\label{fig:flux_surfaces_W7-X_LHD}
\end{figure}

Analytical and numerical results along the paper will be illustrated by calculations in W7-X and LHD configurations. In figure \ref{fig:flux_surfaces_W7-X_LHD}, we show the flux surface of each device on which these calculations will be done.

\begin{figure}[h!]
	\centering
	\begin{subfigure}{0.45\linewidth}
		\includegraphics[trim={-2mm 0mm 0mm 0mm}] {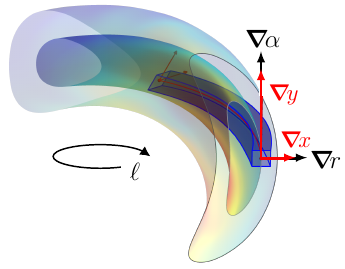}
	\end{subfigure}%
	\begin{subfigure}{0.59\linewidth}
		\includegraphics[trim={-4mm -3mm 0mm 0mm}] {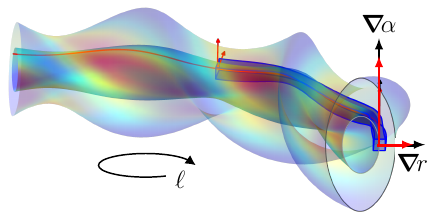}
	\end{subfigure} 
	\caption{Representation of flux tubes like the ones employed in the gyrokinetic simulations of this paper in the standard configuration of W7-X (left) and an inward-shifted configuration of LHD (right).}
\label{fig:flux_tubes_W7-X_LHD}
\end{figure}

Gyrokinetic simulations are carried out using the flux-tube code \texttt{stella}. Flux tubes like the ones that we will employ are represented in figure \ref{fig:flux_tubes_W7-X_LHD}. Details on the implementation of the equations and conventions used in \texttt{stella} can be found in \cite{Barnes_jcp_2019, GonzalezJerez_jpp_2022}. The domain in $\ell$ (i.e.~the length of the flux tube) of the simulations in the paper corresponds to three poloidal turns for linear simulations and approximately one  poloidal  turn for nonlinear simulations. When giving results of numerical simulations, we will use the more common convention $k_x, k_y$ instead of $k_r, k_\alpha$, where $k_x = k_r$ and $k_y = k_\alpha / r$.

The nonlinear \texttt{stella} simulations are performed with a resolution of $(N_x, N_y, N_\mu, N_{u}) = (91, 91, 12, 48)$ grid points, and $N_{\ell}=49$ for W7-X whereas $N_{\ell}=97$ for LHD. The perpendicular box size in the directions $x$ and $y$ is $94\rho_i \times 94\rho_i$, which corresponds to a box in Fourier space such that $(k_{x,{\rm max}}\rho_i, k_{y,{\rm max}}\rho_i) = (2,2)$. Unless stated otherwise, for linear simulations we take $(N_\ell, N_\mu, N_u) = (513, 12, 96)$ and $k_x = 0$.

\begin{figure}[h!]
	\centering  
	\includegraphics{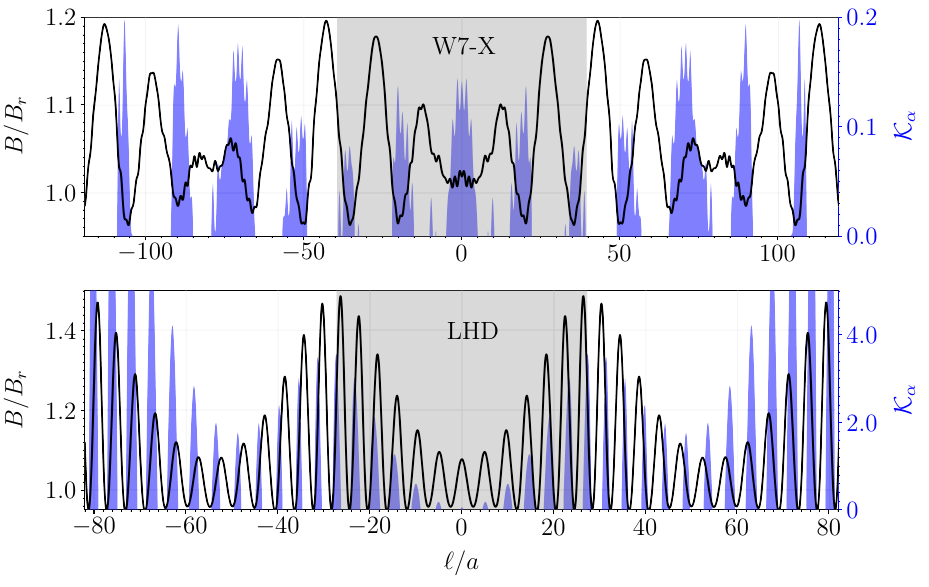} 
	\vspace{-8mm}
	\caption{Magnetic field strength and bad curvature regions along three  poloidal turns of the flux tubes represented in figure \ref{fig:flux_tubes_W7-X_LHD}. The gray background corresponds to one  poloidal turn. Bad curvature regions for $k_r = 0$ are those for which $\mathcal{K}_\alpha > 0$, with $\mathcal{K}_\alpha = (a^2 B_r / B^3) (\mathbf{B} \times \nablammm B) \cdot \nablam \alpha$. Here, $B_r = 2\Psi_t(a)/a^2$ is a reference value for the magnetic field strength.}
	\label{fig:geometry_W7-X_LHD}
\end{figure}

In figure \ref{fig:geometry_W7-X_LHD} we represent some relevant geometric quantities along the simulated flux tubes, centered at the outboard midplane of the bean-shaped cross-section of W7-X, and analogously for LHD, as illustrated in figure \ref{fig:flux_tubes_W7-X_LHD}.

\section{Localization of ITG modes along magnetic field lines}
\label{sec:local_models}

In relevant regions of parameter space, ITG modes are sufficiently localized along the field line so that versions of the gyrokinetic equations that are local in $\ell$ can give reasonably accurate predictions of the growth rate and frequency of the modes. The goal of this section and accompanying appendices is not to provide a thorough discussion on the localization of ITG modes (see, for example, \cite{Plunk2014, Zocco2016, Zocco2018, Rodriguez2025} for detailed discussions with emphasis on stellarator geometry), but to justify, at least partially, the use of equations  local in $\ell$ \mbox{ for} the analytical calculation of section \ref{sec:solution_ITG_dispersion_relation}.

Consider equation \eq{eq:h_equation_local_linear} and assume that  the term $ - \omega_{ds} 
\check h_s$ dominates over the parallel streaming term $i (u \partial_\ell - \mu \partial_\ell B\partial_u) \check h_s$. Neglecting parallel streaming, equation \eq{eq:h_equation_local_linear} becomes
\begin{equation}\label{eq:h_equation_local_linear_toroidal}
\left(\omega - \omega_{ds} \right)
\check h_s
=
\left(\omega -  \omega_{*s}^T  \right)
\frac{Z_s e}{T_s} \check \varphi J_0(k_\perp \rho_s) f_{Ms},
\end{equation}
which, together with \eq{eq:quasineutrality_h_linear}, are the equations for the so-called toroidal branch of the ITG mode. Equation \eq{eq:h_equation_local_linear_toroidal} for the toroidal instability is local in the coordinate $\ell$, in the sense that each point along the field line evolves in time independently. Hence, one has a growth rate $\gamma(\ell)$ for each value of $\ell$. Physically, \eq{eq:h_equation_local_linear_toroidal} describes the actual instability only if the exact ITG mode is strongly localized around a point (or a discrete number of points) along the field line. If that is the case, the maximum of $\{\gamma(\ell), \, \ell \in [0,\ell_{\rm max}] \}$ gives the local approximation to the actual growth rate of the complete equation \eq{eq:h_equation_local_linear}. In \ref{sec:accuracy_toroidal_ITG_equations}, by means of examples, we show that, in certain regions of parameter space, the toroidal ITG equations can predict well the localization of the modes and the value of the real frequency, although their prediction of the value of the growth rate is, in general, not very accurate.

Instead of neglecting the parallel streaming term, one can obtain more refined local equations by modelling the effect of the parallel streaming through a finite, constant parallel wavenumber $k_{||}$. That is,
\begin{equation}\label{eq:h_equation_local_linear_finite_k_par}
\fl
- u k_{||} \check h_s
+
\left(\omega - \omega_{ds} \right)
\check h_s
=
\left(\omega -  \omega_{*s}^T  \right)
\frac{Z_s e}{T_s} \check \varphi J_0(k_\perp \rho_s) f_{Ms}.
\end{equation}
Obviously, equation \eq{eq:h_equation_local_linear_finite_k_par} for $k_{||} = 0$ reduces to the toroidal ITG equation \eq{eq:h_equation_local_linear_toroidal}. In order to calculate $\omega$ in this model, one has to solve the local dispersion relation determined by simultaneously imposing \eq{eq:h_equation_local_linear_finite_k_par} and the quasineutrality equation \eq{eq:quasineutrality_h_linear}. An explicit expression for the local dispersion relation can be derived~\cite{Zocco2018, Parisi2020} that, however, involves complicated integrals that, in general, must be evaluated numerically. We give this expression in equations \eq{eq:dispersion_relation} and \eq{eq:result_Da} in \ref{sec:local_dispersion_relation}. Employing the integration path on the complex plane proposed in \cite{Parisi2020} to efficiently compute the integral in \eq{eq:result_Da}, we have written a \texttt{python} script that solves the local dispersion relation fast and accurately. In \ref{sec:benchmark_python_script} we check that, for $k_{||} = 0$, the \texttt{python} script that solves the dispersion relation defined by \eq{eq:dispersion_relation} and \eq{eq:result_Da} gives the same values for the growth rate and the real frequency as a \texttt{stella} simulation without parallel streaming terms.

In \ref{sec:accuracy_model_finite_k_par} we argue that in regions of parameter space where the mode is localized, one can meaningfully select a value of $k_{||}$, extracted from complete linear gyrokinetic simulations, so that both the actual real frequency and growth rate of the mode are approximately matched. This exercise has no particular predictive power (because one needs to run complete gyrokinetic simulations to fix $k_{||}$), but illustrates the fact that the toroidal ITG equations or refinements thereof can be reasonable approximations to the exact equations. 

As explained at the beginning of this section, the above brief discussion on mode localization is mainly intended to justify, at least partially, the use of local equations for the analytical calculation of section \ref{sec:solution_ITG_dispersion_relation}.

\section{Analytical expression for the modification of the toroidal ITG growth rate by impurities}
\label{sec:solution_ITG_dispersion_relation}

Equation \eq{eq:dispersion_relation}, with the definition of $D_a$, $a = i, z$ given in \eq{eq:result_Da}, is the local ITG dispersion relation in a plasma consisting of bulk ions, electrons and a species of impurity ions. We proceed to find an analytical expression for the change of the growth rate due to impurities. Specifically, we would like to compute
\begin{equation}
\overline{\Delta\gamma}
=
\frac{\gamma - \gamma_0}{\gamma_0}
,
\end{equation}
the normalized difference between the growth rate with impurities, $\gamma$, and the growth rate without impurities, which we denote by $\gamma_0$. 

In order to make analytical progress, we need to simplify the local ITG dispersion relation. We assume $k_{||} = 0$ (i.e.~we consider the toroidal branch of the ITG mode) and small Larmor radius. We also assume large $\eta_i$ (i.e. bulk ions very far from marginality), highly-charged impurities and small impurity concentration.

In \ref{sec:dispersion_relation_small_krho}, we take the $k_{||} = 0$, small Larmor radius limit of the local dispersion relation \eq{eq:dispersion_relation}. This limit of the dispersion relation is given in equation \eq{eq:quasineutrality_h_linear_ITG_with_impurity_ions_new_coor} with the expressions for $I_i$ and $I_z$ given by \eq{eq:integral_I_final}. The large $\eta_i$ limit and the assumptions of highly-charged impurities and small impurity concentration simplify $I_i$ and $I_z$, allowing us to solve the dispersion relation perturbatively. In subsection \ref{sec:fluid_limit_toroidal_ITG}, we give the large $\eta_i$ limit of $I_i$. In subsection \ref{sec:analytical_solution_toroidal_ITG}, we employ the assumptions of highly-charged impurities and small impurity concentration to simplify $I_z$, solve the dispersion relation perturbatively and finally give the analytical expression for $\overline{\Delta\gamma}$.

\subsection{Large $\eta_i$ limit of $I_i$}
\label{sec:fluid_limit_toroidal_ITG}

Here, the large $\eta_i$ limit is defined by the orderings
\begin{equation}\label{eq:ordering_large_etai_1}
\frac{\omega_{*i}}{\omega} \sim 1,
\end{equation}
\begin{equation}\label{eq:ordering_large_etai_2}
\frac{\omega_{di}}{\omega} \sim \frac{1}{\eta_i} \ll 1.
\end{equation}

In order to calculate the large $\eta_i$ limit of $I_i$, we need the expansion \eq{eq:expansion_Z_large_zeta_Imzeta_positive} of the plasma dispersion function and the definition of $\Omega_i$ given in \eq{eq:def_Omega_a}. Assuming that $|\Omega_i^{1/2}| \gg 1$ and that ${\rm Im}(\Omega_i^{1/2}) > 0$ is not exponentially small in $\eta_i^{-1} \ll 1$\footnote{These assumptions can be checked \emph{a posteriori}.}, we can write
\begin{equation}\label{eq:expansion_Z_large_zeta_without_exponential}
{\cal Z}(\Omega_i^{1/2}) = -\frac{1}{\Omega_i^{1/2}}\left(
1 + \frac{1}{2\Omega_i} + \frac{3}{4}\frac{1}{\Omega_i^2} + \dots
\right)
\end{equation}
and
\begin{equation}\label{eq:expansion_Z2_large_zeta_without_exponential}
{\cal Z}^2(\Omega_i^{1/2}) = \frac{1}{\Omega_i}\left(
1 + \frac{1}{\Omega_i} + \frac{7}{4}\frac{1}{\Omega_i^2} + \dots
\right).
\end{equation}

Using \eq{eq:expansion_Z_large_zeta_without_exponential} and \eq{eq:expansion_Z2_large_zeta_without_exponential}, we easily find the lowest order contributions to $I_i$ (given in \eq{eq:integral_I_final}) in the large $\eta_i$ expansion defined by \eq{eq:ordering_large_etai_1} and \eq{eq:ordering_large_etai_2},
\begin{equation}\label{eq:Ii_large_etai}
I_i = 1 - \frac{\Omega_{*i}}{\Omega_i} - \eta_i \frac{\Omega_{*i}}{\Omega_i^2},
\end{equation}
where the definition of $\Omega_{*i}$ is given in \eq{eq:def_Omega_stara}.

\subsection{Solution to the dispersion relation for small impurity concentration}
\label{sec:analytical_solution_toroidal_ITG}

The dispersion relation \eq{eq:quasineutrality_h_linear_ITG_with_impurity_ions_new_coor} with $I_i$ given by \eq{eq:Ii_large_etai} reads
\begin{eqnarray}\label{eq:dispersion_relation_large_etai}
\fl
\left(
1 - \frac{\omega_{*i}}{\omega} - \eta_i\omega_{di,0} \frac{\omega_{*i}}{\omega^2}
\right)
+
\frac{Z_z^2 n_z T_i}{Z_i^2 n_i T_z}
I_z
-
\left(
1
+ \frac{T_i}{Z_i T_e}
+ \frac{Z_z^2 n_z T_i}{Z_i^2 n_i T_z} 
+ \frac{Z_z n_z T_i}{Z_i^2 n_i T_e}
\right)
=
0,
\end{eqnarray}
where we have used $n_e = Z_i n_i + Z_z n_z$ to eliminate the electron density from the equation. For the moment, we take $I_z$ as given in \eq{eq:integral_I_final} for $a = z$; that is,
\begin{eqnarray}\label{eq:Iz}
\fl
I_z(\Omega_z)
=
\left\{
\Omega_z - \Omega_{*z}\left[ 1+ \eta_z \left(2\Omega_z - 1\right)\right]
\right\}
{\cal Z}^2(\Omega_z^{1/2})
-2 \Omega_{*z} \eta_z \Omega_z^{1/2}{\cal Z}(\Omega_z^{1/2}).
\end{eqnarray}

In order to find analytical solutions to the dispersion relation, we assume highly-charged impurities,
\begin{equation}\label{eq:Z_z_large}
\frac{1}{Z_z} \ll 1,
\end{equation}
and small impurity concentration,
\begin{equation}\label{eq:Z_eff_small}
\varepsilon := \frac{Z_z^2 n_z}{Z_i^2 n_i} \ll 1.
\end{equation}
Note that, for $Z_i = 1$, one has
\begin{equation}\label{eq:relation_varepsilon_Zeff}
\varepsilon \simeq Z_{\rm eff} -1,
\end{equation}
where $Z_{\rm eff} = (Z_i^2 n_i + Z_z^2 n_z) / (Z_i n_i + Z_z n_z)$ is the usual definition of the effective charge in a plasma with one impurity species.

We turn to expand \eq{eq:dispersion_relation_large_etai} in $\varepsilon \ll 1$ and solve the dispersion relation perturbatively. The lowest-order of the expansion corresponds to the case in which there are no impurities; i.e.~$\varepsilon = 0$. We expand the frequency as $\omega = \omega_0 + \omega_1 + \dots$, where $\omega_0$ is the solution of \eq{eq:dispersion_relation_large_etai} for $\varepsilon = 0$ and $\omega_1$ gives the correction to $\omega_0$ that is linear in $\varepsilon $. We will see below that $\omega_1 = O(Z_z^{-1}\varepsilon \omega_0)$.

For $\varepsilon = 0$, equation \eq{eq:dispersion_relation_large_etai} gives
\begin{eqnarray}\label{eq:dispersion_relation_large_etai_approx_lowest_order}
\frac{\omega_{*i}}{\omega_0} + \eta_i\omega_{di,0} \frac{\omega_{*i}}{\omega_0^2}
+
\frac{T_i}{Z_i T_e}
=
0,
\end{eqnarray}
whose solutions are
\begin{equation}\label{eq:omega0_ITG_with_impurities_aux}
\omega_{0\pm}
=
\frac{\omega_{*e}}{2}
\left(
1
\pm
\sqrt{1
+ \frac{4 \eta_i \omega_{di, 0}}{\omega_{*e}}} \,
\right),
\end{equation}
where we have used that $\omega_{*e} \simeq - ({Z_i T_e}/{T_i}) \omega_{*i}$ to write the expression for $\omega_{0\pm}$ in a slightly more compact way. A necessary condition for \eq{eq:omega0_ITG_with_impurities_aux} to give an instability is that its right-hand side has a non-zero imaginary part. This happens if
\begin{equation}\label{eq:def_bad_curvature}
\eta_i  \omega_{*i}\omega_{di, 0} > 0,
\end{equation}
that defines the bad curvature regions. In what follows, we assume \eq{eq:def_bad_curvature} and that the lowest-order frequencies $\omega_{0\pm}$ are determined by evaluating the right-hand side of \eq{eq:omega0_ITG_with_impurities_aux} at a point of bad curvature (we are assuming that $\omega_{di, 0}$  and $\omega_{dz, 0}$ are positive; this assumption is made in the course of the derivation presented in \ref{sec:dispersion_relation_small_krho}). Note that $\Omega_{i0+} := \omega_{0+}/\omega_{di,0}$ and $\Omega_{i0-} := \omega_{0-}/\omega_{di,0}$ satisfy the assumptions made before \eq{eq:expansion_Z_large_zeta_without_exponential}.

From now on, we assume that
\begin{equation}
1
+ \frac{4 \eta_i \omega_{di, 0}}{\omega_{*e}} < 0,
\end{equation}
a sufficient condition for \eq{eq:omega0_ITG_with_impurities_aux} to give an instability. Then, the lowest-order frequencies are complex and read
\begin{equation}\label{eq:omega0_complex}
\omega_{0\pm}
=
\frac{\omega_{*e}}{2}
\left(
1
\pm
i
\sqrt{
-\frac{4 \eta_i \omega_{di, 0}}{\omega_{*e}}-1} \,
\right).
\end{equation}
If $\omega_{*e} > 0$ (resp. $\omega_{*e} < 0$), the mode with frequency $\omega_{0+}$ (resp. $\omega_{0-}$) is unstable. Let us work out the corrections $\omega_{1\pm}$ to the lowest-order frequencies \eq{eq:omega0_complex}; that is, let us study perturbatively how impurities modify an unstable ITG mode.

The expansion of \eq{eq:dispersion_relation_large_etai} to next order in $\varepsilon \ll 1$ gives
\begin{eqnarray}\label{eq:dispersion_relation_large_etai_approx_next_to_lowest_order}
\fl
\left(
1 + \frac{2\eta_i \omega_{di,0}}{\omega_{0\pm}}
\right) \frac{\omega_{*i}}{\omega_{0\pm}^2}\omega_{1\pm}
+
\frac{Z_z^2 n_z T_i}{Z_i^2 n_i T_z}
\left(
I_z (\omega_{0\pm}/\omega_{dz,0})-1 - \frac{T_z}{Z_z T_e}
\right)
=
0.
\end{eqnarray}
Hence, the correction to the lowest-order frequency is
\begin{eqnarray}\label{eq:omega1}
\fl
 \omega_{1\pm}
=
-
\frac{Z_z^2 n_z T_i}{Z_i^2 n_i T_z}
 \frac{\omega_{0\pm}^2}{\omega_{*i}}
\left(
1 + \frac{2\eta_i \omega_{di,0}}{\omega_{0\pm}}
\right)^{-1}
\left(
I_z (\omega_{0\pm}/\omega_{dz,0})-1 - \frac{T_z}{Z_z T_e}
\right)
.
\end{eqnarray}
At first sight, one might think that the last term on the left-hand side of this expression is negligible because it is small in $1/Z_z \ll 1$, but we will see below that it is of the same order as $I_z (\omega_{0\pm}/\omega_{dz,0})-1$.

Let us make the expression for $\omega_{1\pm}$ more explicit by using that $|{\omega_{0\pm}}/{\omega_{dz,0}}| \sim Z_z |{\omega_{0\pm}}/{\omega_{di,0}}| \gg 1$. For $\eta_z$, we take the maximal ordering $\eta_z \sim |{\omega_{0\pm}}/{\omega_{dz,0}}|$, which allows us to study the effect of both the impurity density and temperature gradients. Employing \eq{eq:expansion_Z_large_zeta_Imzeta_positive}, we expand $I_z(\omega_{0\pm}/\omega_{dz,0})$ to lowest order,
\begin{equation}\label{eq:Iz_including_etaz_lowest_order}
I_z(\omega_{0\pm}/\omega_{dz,0}) = 1 - \frac{\omega_{*z}}{\omega_{0\pm}} - \eta_z \frac{\omega_{*z} \omega_{dz,0}}{\omega_{0\pm}^2} + \dots
\end{equation}
Plugging this result in \eq{eq:omega1}, we have
\begin{eqnarray}\label{eq:omega1_with_Tzprime}
\fl
 \omega_{1\pm}
=
\frac{Z_z^2 n_z T_i}{Z_i^2 n_i T_z}
 \frac{\omega_{*z}}{\omega_{*i}}
\left(
1 + \frac{2\eta_i \omega_{di,0}}{\omega_{0\pm}}
\right)^{-1}
\left(
  \omega_{0\pm} + \eta_z  \omega_{dz,0}
   + \frac{T_z}{Z_z T_e} \frac{\omega_{0\pm}^2}{\omega_{*z}}
\right)
.
\end{eqnarray}
Noting that
\begin{equation}
1 + \frac{2\eta_i \omega_{di,0}}{\omega_{0\pm}}
= \pm i \sqrt{
-\frac{4 \eta_i \omega_{di, 0}}{\omega_{*e}}-1} \, ,
\end{equation}
we obtain
\begin{eqnarray}\label{eq:Imomega1_with_Tzprime}
\fl
&&
{\rm Im} (\omega_{1\pm})
=
\mp
\frac{Z_z^2 n_z T_i}{Z_i^2 n_i T_z}
 \frac{\omega_{*z}}{\omega_{*i}}
\left(
-\frac{4 \eta_i \omega_{di, 0}}{\omega_{*e}}-1
\right)^{-1/2}
\left(
{\rm Re} (\omega_{0\pm}) + \eta_z  \omega_{dz,0} + \frac{T_z}{Z_z T_e} \frac{{\rm Re}(\omega_{0\pm}^2)}{\omega_{*z}}
\right) =
\nonumber
\\[5pt]
\fl
&&
\hspace{0cm}
\mp
\frac{Z_z^2 n_z T_i}{Z_i^2 n_i T_z}
 \frac{\omega_{*z}}{\omega_{*i}}
 \frac{\omega_{*e}}{2}
\left(
-\frac{4 \eta_i \omega_{di, 0}}{\omega_{*e}}-1
\right)^{-1/2}
\left[
1 + \eta_z \frac{2 \omega_{dz,0}}{\omega_{*e}}
+ \frac{T_z}{Z_z T_e} \frac{\omega_{*e}}{\omega_{*z}}
 \left(
 1 + \frac{2 \eta_i \omega_{di, 0}}{\omega_{*e}} 
\right)
\right],
\end{eqnarray}
where we have used
\begin{equation}
{\rm Re}(\omega_{0\pm})=
\frac{\omega_{*e}}{2}
\end{equation}
and
\begin{equation}
{\rm Re}(\omega_{0\pm}^2)=
\frac{\omega_{*e}^2}{2}\left(
1 + \frac{2 \eta_i \omega_{di, 0}}{\omega_{*e}}
\right).
\end{equation}

Let us focus on the unstable mode, whose growth rate we call $\gamma_0$. Note that $\gamma_0 =  {\rm Im} (\omega_{0+})$ if $\omega_{*e} > 0$ and $\gamma_0 =  {\rm Im} (\omega_{0-})$ if $\omega_{*e} < 0$. We denote the modification of this growth rate by the impurities by $\gamma_1$, where $\gamma_1 = {\rm Im} (\omega_{1+})$ if $\omega_{*e} > 0$ and $\gamma_1 = {\rm Im} (\omega_{1-})$ if $\omega_{*e} < 0$. Then,
\begin{eqnarray}\label{eq:gamma1}
\fl
&&
\overline{\Delta\gamma} \simeq \frac{\gamma_1}{\gamma_0}
=
\nonumber
\\[5pt]
\fl
&&\hspace{0.5cm}
-
\frac{Z_z^2 n_z T_i}{Z_i^2 n_i T_z}
 \frac{\omega_{*z}}{\omega_{*i}}
\left(
-\frac{4 \eta_i \omega_{di, 0}}{\omega_{*e}}-1
\right)^{-1}
\left[
1 + \eta_z \frac{2 \omega_{dz,0}}{\omega_{*e}}
+ \frac{T_z}{Z_z T_e} \frac{\omega_{*e}}{\omega_{*z}}  \left(
1 + \frac{2 \eta_i \omega_{di, 0}}{\omega_{*e}} 
\right)
\right].
\end{eqnarray}
Finally, using $\omega_{*e} \simeq - ({Z_i T_e}/{T_i}) \omega_{*i}$, we can write $\overline{\Delta\gamma} $ in terms of ionic quantities,
\begin{eqnarray}\label{eq:gamma_final}
\fl
&&
\overline{\Delta\gamma}
\simeq
-
\frac{Z_z^2 n_z T_i}{Z_i^2 n_i T_z}
\left(
\frac{4 T_i \eta_i \omega_{di, 0}}{Z_i T_e \omega_{*i}}-1
\right)^{-1}
\left[
 \frac{\omega_{*z}}{\omega_{*i}} - \frac{2 T_i}{Z_i T_e }  \frac{\eta_z \omega_{*z} \omega_{dz,0}}{\omega_{*i}^2}  
+ \frac{Z_i T_z}{Z_z T_i}  \left(
\frac{2 T_i \eta_i \omega_{di, 0}}{Z_i T_e \omega_{*i}} - 1 
\right)
\right].
\nonumber
\\
\fl
&&
\end{eqnarray}
The relevant question is whether this expression for $\overline{\Delta\gamma}$, although obtained under crude assumptions, captures well the sizes, scalings and signs of the different physical effects involved in the modification of the ITG growth rate by impurities. We discuss this in section \ref{sec:analytical_expression_vs_linear_simulations}.

\section{Modification of the ITG growth rate by impurities: comparison between analytical predictions and  linear gyrokinetic simulations}
\label{sec:analytical_expression_vs_linear_simulations}

Instead of focusing on the details of expression \eq{eq:gamma_final}, let us discuss its general structure. Expression \eq{eq:gamma_final} can be written as
\begin{eqnarray}\label{eq:Deltagamma_structure}
\overline{\Delta\gamma} 
=
\frac{\varepsilon}{Z_z}
\left(
- C_n \frac{a}{L_{n_z}}
+ C_T \frac{1}{Z_z}\frac{a}{L_{T_z}}
-
C_0
\right),
\end{eqnarray}
where the coefficients $C_n$, $C_T$ and $C_0$ are independent of the impurity profile gradients and, for $k_x = 0$, they are independent of $k_y$. Note that ${\rm sign}(C_n) = {\rm sign}(a/L_{n_i})$, $C_0$ is positive for sufficiently large $\eta_i$ and ${\rm sign}(C_T) = {\rm sign}(a/L_{T_i})$. To prove the last property, one has to use \eq{eq:def_bad_curvature}. Note also that $\overline{\Delta\gamma}$ scales with $\varepsilon/Z_z$, that sets the typical size (advanced a few lines after \eq{eq:relation_varepsilon_Zeff}) of the modification of the ITG growth rate by impurities. The expression of $\overline{\Delta\gamma}$ is the sum of three terms corresponding to three physically different effects: a term associated to the impurity density gradient, a term associated to the impurity temperature gradient and a term that does not depend on the gradients of the impurity profiles that corresponds to dilution. Considering the signs of the coefficients $C_n$, $C_T$ and $C_0$, expression \eq{eq:Deltagamma_structure} predicts a stabilizing (resp.~destabilizing) effect of the impurity density gradient if $L_{n_i}/L_{n_z} > 0$ (resp. $L_{n_i}/L_{n_z} < 0$), a destabilizing (resp.~stabilizing) effect of the impurity temperature gradient if $L_{T_i}/L_{T_z} > 0$ (resp. $L_{T_i}/L_{T_z} < 0$), and a stabilizing effect from dilution. Observe that the effect of the impurity temperature gradient is small in $1/Z_z \ll 1$.

\begin{figure}[h!]
\centering
\hspace*{-5mm}
\includegraphics{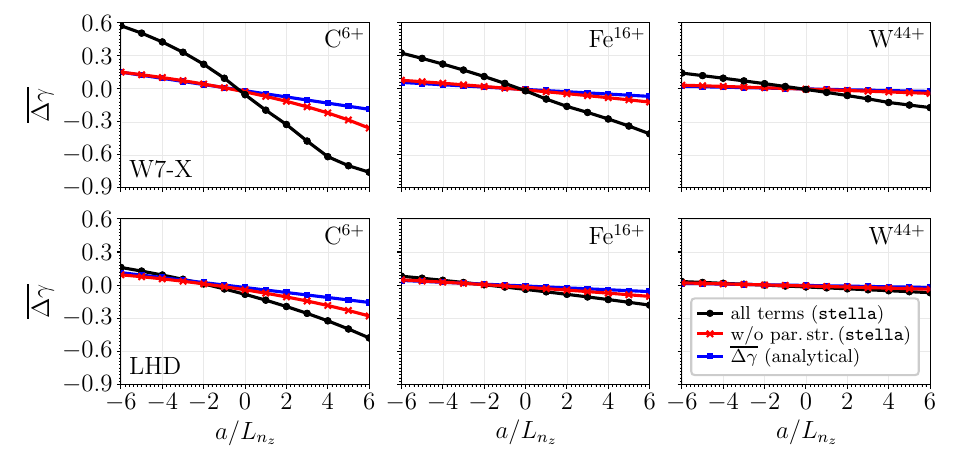}
\vspace{-9mm}
\caption{$\overline{\Delta\gamma}$ versus $a/L_{n_z}$ for different impurities obtained from complete linear gyrokinetic simulations (black), the exact solution of the toroidal ITG linear gyrokinetic equation (red) and the analytical approximation to the solution of the toroidal ITG dispersion relation (blue). Here, $a/L_{T_z}=0$ and $n_z/n_i$ is chosen so that $\varepsilon = Z_z^2n_z / (Z_i^2 n_i) = 0.4$.}
\label{fig:Deltagamma_vs_density_gradient}
\end{figure}

\begin{figure}[h!]
\centering
\hspace*{-5mm}
\includegraphics{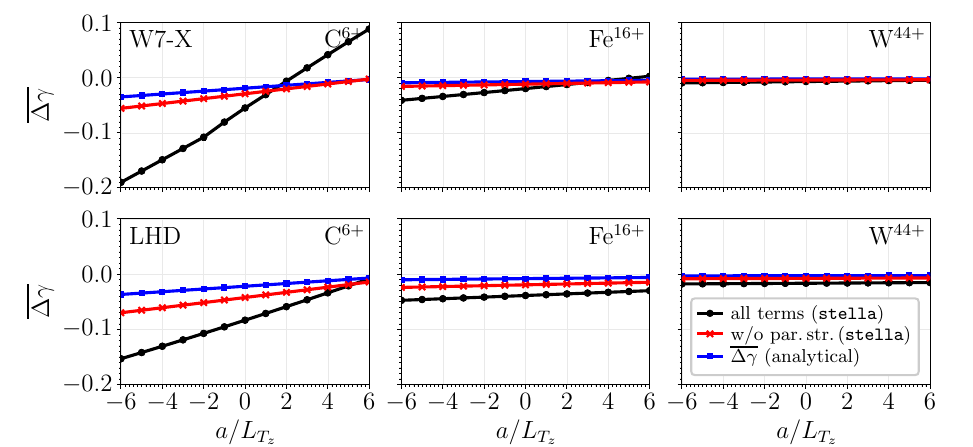}
\vspace{-7mm}
\caption{$\overline{\Delta\gamma}$ versus $a/L_{T_z}$ for different impurities obtained from complete linear gyrokinetic simulations (black), the exact solution of the toroidal ITG linear gyrokinetic equation (red) and the analytical approximation to the solution of the toroidal ITG dispersion relation (blue). Here, $a/L_{n_z}=0$ and $n_z/n_i$ is chosen so that $\varepsilon = Z_z^2n_z / (Z_i^2 n_i) = 0.4$.}
\label{fig:Deltagamma_vs_temperature_gradient}
\end{figure}

\begin{figure}[h!]
\centering
\hspace*{-2mm}
\includegraphics{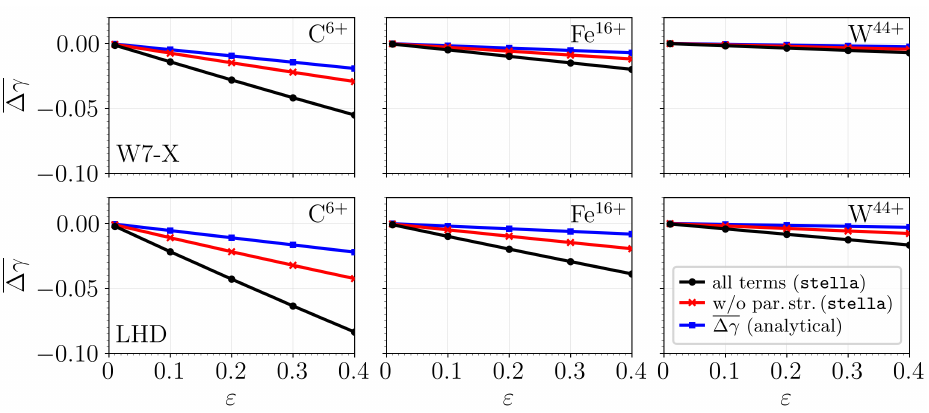}
\vspace{-8mm}
\caption{$\overline{\Delta\gamma}$ versus $\varepsilon$ for different impurities obtained from complete linear gyrokinetic simulations (black), the exact solution of the toroidal ITG linear gyrokinetic equation (red) and the analytical approximation to the solution of the toroidal ITG dispersion relation (blue). Here, $a/L_{n_z} = a/L_{T_z} = 0$ and  $\varepsilon$ is varied by varying $n_z/n_i$.}
\label{fig:Deltagamma_vs_varepsilon}
\end{figure}

\begin{figure}[h!]
\centering
\hspace*{-2mm}
\includegraphics{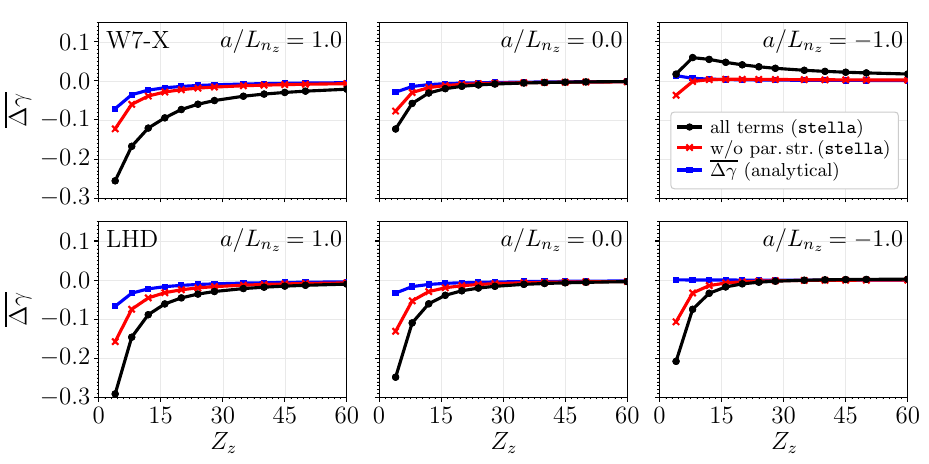}
\vspace{-9mm}
\caption{$\overline{\Delta\gamma}$ versus $Z_z$ obtained from complete linear gyrokinetic simulations (black), the exact solution of the toroidal ITG linear gyrokinetic equation (red) and the analytical approximation to the solution of the toroidal ITG dispersion relation (blue). Here, $m_z$ is the mass of carbon and $a/L_{T_z}=0$, whereas $a/L_{n_z}=1$ (left column), $a/L_{n_z}=0$ (middle column) and $a/L_{n_z}=-1$ (right column).  The impurity concentration, $n_z/n_i$, is chosen so that $\varepsilon = Z_z^2n_z / (Z_i^2 n_i) = 0.4$.}
\label{fig:Deltagamma_vs_Zz}
\end{figure}

The typical sizes and main scalings of the different physical effects involved in the modification of the ITG growth rate by impurities are well captured by \eq{eq:gamma_final}  (or \eq{eq:Deltagamma_structure}). This is the case even for realistic values of $Z_{\rm eff}$. We show this in figures \ref{fig:Deltagamma_vs_density_gradient}, \ref{fig:Deltagamma_vs_temperature_gradient}, \ref{fig:Deltagamma_vs_varepsilon} and \ref{fig:Deltagamma_vs_Zz}, where we give calculations for W7-X and LHD. From here on and in the rest of the paper, in the numerical examples we consider hydrogen bulk ions with $a/L_{T_i} = 3$, $a/L_{n_i} = 1$, and we take $T_z = T_i = T_e$.

In figure \ref{fig:Deltagamma_vs_density_gradient} we show $\overline{\Delta\gamma}$ versus $a/L_{n_z}$ at $a/L_{T_z} = 0$ for carbon, iron and tungsten. The black curves are full linear simulations, the red curves are exact values of the toroidal ITG growth rate calculated from \texttt{stella} simulations where the parallel streaming terms have been switched off and the blue curves are obtained by evaluating \eq{eq:gamma_final}. We have taken $\varepsilon = 0.4$, which corresponds to $Z_{\rm eff} \simeq 1.4$ (observe that the exact value of $Z_{\rm eff}$ depends slightly on the specific impurity under consideration). The agreement is better for larger $Z_z$, but even for carbon the main predictions of the analytical formula work well. In figure  \ref{fig:Deltagamma_vs_temperature_gradient} we show analogous calculations for $\overline{\Delta\gamma}$ versus $a/L_{T_z}$ at $a/L_{n_z} = 0$. In figure \ref{fig:Deltagamma_vs_varepsilon} we represent $\overline{\Delta\gamma}$ as a function of $\varepsilon$ at $a/L_{n_z} = a/L_{T_z} = 0$. Lastly, in figure \ref{fig:Deltagamma_vs_Zz} we give $\overline{\Delta\gamma}$ as a function of $1/Z_z$ for $ a/L_{T_z} = 0$ at different values of $a/L_{n_z}$. The scaling with $1/Z_z$ predicted by the analytical calculation is nicely verified. In \texttt{stella} simulations included in these figures, we have taken $k_y \rho_i = 0.5$.

\FloatBarrier

\section{Correlation between the analytical expression for the modification of the ITG growth rate by impurities and the ion heat flux obtained from nonlinear gyrokinetic simulations}
\label{sec:analytical_expression_vs_nonlinear_simulations}

\begin{figure}[h!]
\centering
\includegraphics{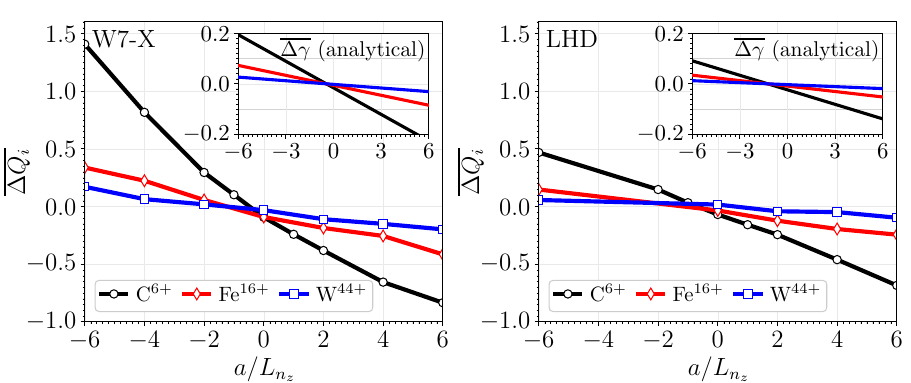}
\vspace{-2mm}
\caption{$\overline{\Delta Q_i}$ obtained from nonlinear \texttt{stella} simulations (assuming adiabatic electrons) versus $a/L_{n_z}$ for several impurity species. In the inset, $\overline{\Delta\gamma}$ calculated analytically, already shown in figure \ref{fig:Deltagamma_vs_density_gradient}, is included. Here, $a/L_{T_z}=0$, whereas $n_z/n_i$ is chosen so that $\varepsilon = Z_z^2n_z / (Z_i^2 n_i) = 0.4$.}
\label{fig:DeltaQi_vs_nzprime_with_inset}
\end{figure}

It is natural to ask how impurities modify the bulk-ion heat flux. We define $\overline{\Delta Q_i} = (Q_i - Q_{i,0})/Q_{i,0}$, where $Q_{i,0}$ is the ion heat flux without impurities. In figure \ref{fig:DeltaQi_vs_nzprime_with_inset},  we show $\overline{\Delta Q_i}$ versus $a/L_{n_z}$ at $\varepsilon = 0.4$ and $a/L_{T_z} = 0$, and include an inset with the plot of $\overline{\Delta\gamma}$ calculated analytically (already shown in figure \ref{fig:Deltagamma_vs_density_gradient}). We see that $\overline{\Delta Q_i}$ and $\overline{\Delta\gamma}$ have the same sign and a similar dependence on $a/L_{n_z}$.

\begin{figure}[h!]
\centering
\includegraphics{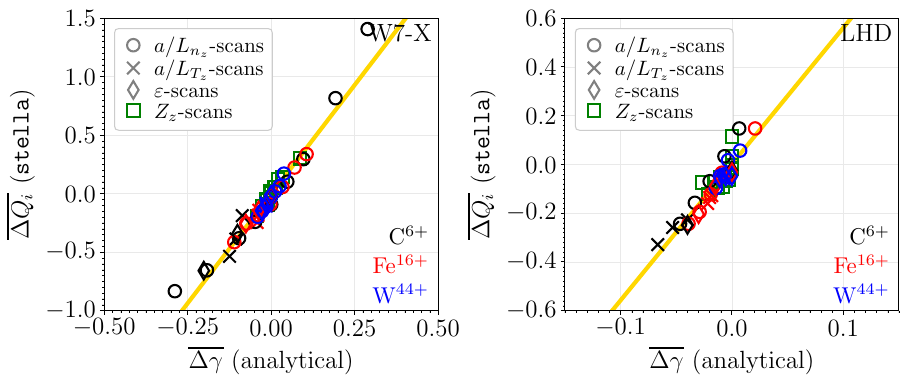}
\vspace{-2mm}
\caption{$\overline{\Delta Q_i}$ obtained from nonlinear \texttt{stella} simulations (assuming adiabatic electrons) versus  $\overline{\Delta\gamma}$ calculated analytically. The details of this scan are given in the text.}
\label{fig:DeltaQi_vs_Deltagamma_ITG}
\end{figure}

Motivated by that similarity, we have performed similar scans for $\overline{\Delta Q_i}$ as those shown in section \ref{sec:analytical_expression_vs_linear_simulations} for $\overline{\Delta\gamma}$, and in figure \ref{fig:DeltaQi_vs_Deltagamma_ITG} we represent $\overline{\Delta Q_i}$ versus  $\overline{\Delta\gamma}$. The ion and impurity parameters considered in these scans are identical to those in section \ref{sec:analytical_expression_vs_linear_simulations}, with a few exceptions. Specifically, to ensure that the impurity effect on $Q_i$ is clearly visible, the $a/L_{T_z}$ scans are carried out with $a/L_{n_z}=2$, the $\varepsilon$ scans with $a/L_{n_z}=4$ and the $Z_z$ scans with $a/L_{n_z}=\{-2, 0, 2\}$. The correlation shown in  figure \ref{fig:DeltaQi_vs_Deltagamma_ITG} is striking, revealing that the expansions in $\varepsilon \ll 1$ and $1/Z_z \ll 1$ work very well even for realistic values of the parameters.

\begin{figure}[h!]
\centering
\includegraphics{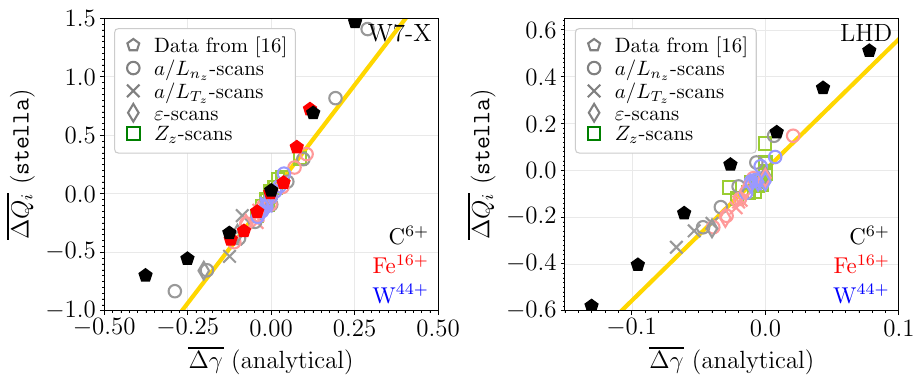}
\vspace{-2mm}
\caption{This figure includes the points from figure \ref{fig:DeltaQi_vs_Deltagamma_ITG}, shown here in fainter colors. In more vivid colors, points corresponding to results in \cite{GarciaRegana2024} have been added. Note that in reference \cite{GarciaRegana2024}, simulations for carbon and iron were carried out in W7-X and for carbon in LHD.}
\label{fig:DeltaQi_vs_Deltagamma_ITG_and_Garcia-Regana_PRL}
\end{figure}

Finally, we explore the correlation between the analytical calculation of  $\overline{\Delta\gamma}$ and $\overline{\Delta Q_i}$ as computed in \cite{GarciaRegana2024}, where electrons are kinetic. In figure \ref{fig:DeltaQi_vs_Deltagamma_ITG_and_Garcia-Regana_PRL}, we add results from \cite{GarciaRegana2024} to the plot shown in figure \ref{fig:DeltaQi_vs_Deltagamma_ITG}. The correlation is still very good for W7-X and only starts to fail for large impurity density gradients (this is expected, because for such large values of the impurity density gradient, the electron density gradient becomes large in the simulations of \cite{GarciaRegana2024} and instabilities different from those driven by the ion temperature gradient might play a relevant role). For LHD, the correlation is excellent for all cases considered.

\section{Conclusions}
\label{sec:conclusions}

We have discussed the impact of impurities on ITG stability and turbulence in stellarators.

In certain asymptotic limits, we have solved the toroidal ITG dispersion relation and obtained a formula for the modification of the ITG growth rate by impurities. The formula is the sum of three terms corresponding to: (i) the effect of the impurity density gradient; (ii) the effect of the impurity temperature gradient; (iii) the effect of dilution (i.e.~of adding impurities with vanishing density and temperature gradients). The analytical calculation predicts when impurities increase or reduce the ITG growth rate. Apart from providing physical insight into the problem, the analytical result predicts the typical size of each effect and the dependence of the modification of the growth rate on fundamental quantities such as the impurity charge, and the impurity density and temperature gradients.

We have also shown that the analytical formula for the modification of the ITG linear growth rate by impurities predicts well the fundamental scalings of the modification of the ITG ion heat flux. What is more, we have shown that the linear ITG calculation captures the main effects on the ion heat flux recently identified in \cite{GarciaRegana2024} even though the nonlinear simulations in \cite{GarciaRegana2024} include kinetic electrons.

The results of this paper, and in particular the clear identification of the different effects by which impurities can modify the turbulent ion heat flux, are expected to find applications in the interpretation of experiments in current devices, and in the design of reactor-relevant operation scenarios with optimized turbulent heat transport.

\ack

This work has been carried out within the framework of the EUROfusion Consortium, funded by the European Union via the Euratom Research and Training Programme (Grant Agreement No 101052200 -- EUROfusion). Views and opinions expressed are however those of the author(s) only and do not necessarily reflect those of the European Union or the European Commission. Neither the European Union nor the European Commission can be held responsible for them. This work was supported by the U.S. Department of Energy under contract number DE-AC02-09CH11466. The United States Government retains a non-exclusive, paid-up, irrevocable, world-wide license to publish or reproduce the published form of this manuscript, or allow others to do so, for United States Government purposes. This research was supported in part by Grants No. PID2021-123175NB-I00 and PID2024-155558OB-I00, funded by Ministerio de Ciencia, Innovaci\'on y Universidades / Agencia Estatal de Investigaci\'on / 10.13039/501100011033 and by ERDF/EU.

\appendix

\section{Accuracy of the model local in $\ell$ for zero parallel wavenumber (toroidal ITG dispersion relation)}
\label{sec:accuracy_toroidal_ITG_equations}

In the numerical examples that follow, we consider hydrogen bulk ions with $a/L_{T_i} = 3$, $a/L_{n_i} = 1$, and we take $T_i = T_e$. We do not include impurities, as they do not change the general discussion on localization.

\begin{figure}[h!]
\centering
\hspace*{-4mm}
\includegraphics{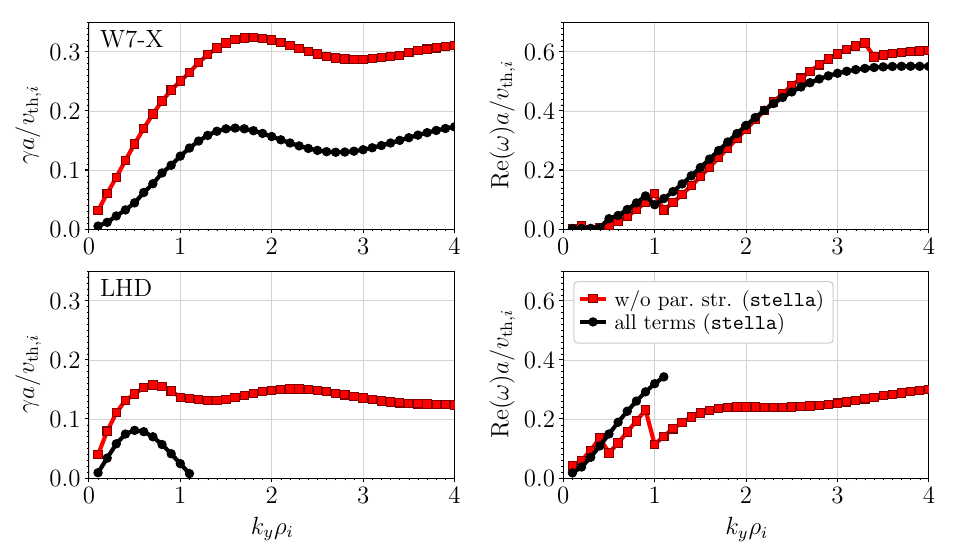}
\vspace{-9mm}
\caption{Growth rate and real frequency in W7-X and LHD as a function of $k_y$. The red curves correspond to the values predicted by the local-in-$\ell$ dispersion relation for the toroidal branch of the ITG mode and the black curves correspond to the exact values.}
\label{fig:examples_accuracy_toroidal_ITG_equations}
\end{figure}

\begin{figure}[h!]
\centering 
\includegraphics{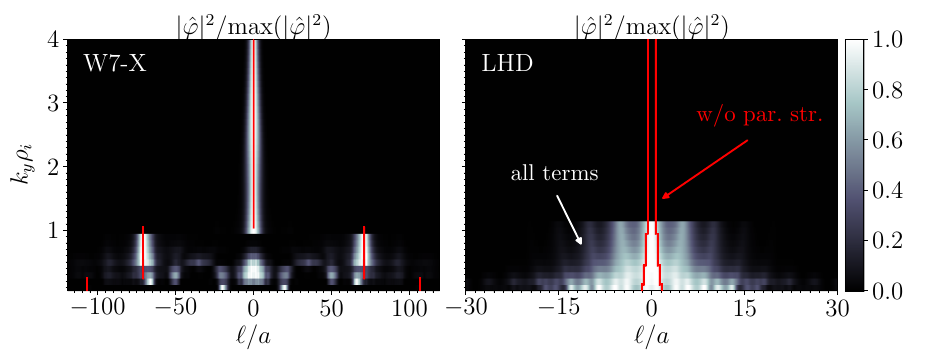}
\vspace{-9mm}
\caption{Parallel mode structure in W7-X and LHD as a function of $k_y$ obtained from complete linear simulations with \texttt{stella}. Here, ${\rm max}(|\hat \varphi |^2)$ indicates the maximum of $|\hat \varphi|^2$ along the simulated flux tube. In red, we indicate points in $\ell$ that give the largest growth rate in the local-in-$\ell$ equations for the toroidal ITG instability. Recall that in LHD the exact ITG equations do not give an instability for $k_y\rho_i \gtrsim 1.2$ (see figure \ref{fig:examples_accuracy_toroidal_ITG_equations}).}
\label{fig:color_map_parallel_structure_varphi}
\end{figure}

In figure \ref{fig:examples_accuracy_toroidal_ITG_equations} we give, as a function of $k_y$, the growth rate and the real frequency predicted by the toroidal branch of the ITG (in red) and compare them with the exact growth rate and frequency obtained from linear simulations with \texttt{stella} (in black). The red curves are calculated by switching off the parallel streaming terms in \texttt{stella}. In general, the local-in-$\ell$ dispersion relation for the toroidal branch of the ITG predicts well the real frequency for a broad range of $k_y$ values whereas the prediction of the growth rate is less accurate. In figure \ref{fig:color_map_parallel_structure_varphi} we show the parallel structure of the modes as a function of $k_y$.

\section{ITG local dispersion relation}
\label{sec:local_dispersion_relation}

We will provide a local dispersion relation for the ITG instability allowing for the presence of impurities. The index $a$ will run over ion species and the index $e$ will denote electrons. From \eq{eq:h_equation_local_linear_finite_k_par}, we deduce
\begin{eqnarray}\label{eq:h_equation_local_linear_ITG_with_impurity_ions}
\check h_a
=
\frac{
\omega -  \omega_{*a}^T
}
{
\omega - u k_{||} - \omega_{da}
}
\,
\frac{Z_a e  \check \varphi}{T_a} J_0(k_\perp \rho_a) f_{Ma},
\end{eqnarray}
We assume a low-$\beta$ magnetohydrodynamic equilibrium, so that
\begin{equation}
\omega_{da} =  \frac{m_a}{Z_a e B^2} (u^2 + \mu B) \bk_\perp\cdot (\bun\times\nabla B),
\end{equation}
and define dimensionless velocity coordinates
\begin{eqnarray}
&& v_{||} = \frac{u}{v_{ta}},
\nonumber\\[5pt]
\hspace{1cm}
&& v_\perp = \sqrt{\frac{2\mu B}{v_{ta}^2}},
\end{eqnarray}
where $v_{ta} = \sqrt{2T_a/m_a}$ is the thermal speed of species $a$. In these coordinates,
\begin{equation}
\omega_{da} = \left( v_{||}^2 + \frac{v_\perp^2}{2} \right) \omega_{da, 0},
\end{equation}
with
\begin{equation}\label{eq:def_omegada0}
\omega_{da, 0} =  \frac{m_a v_{ta}^2}{Z_a e B^2} \bk_\perp\cdot (\bun\times\nabla B).
\end{equation}
If we now define
\begin{equation}
K_{||} = \frac{v_{ta} k_{||}}{|\omega_{da,0}|},
\end{equation}
\begin{equation}\label{eq:def_Omega_a}
\Omega_a = \frac{\omega}{|\omega_{da,0}|},
\end{equation}
\begin{equation}\label{eq:def_Omega_stara}
\Omega_{*a}  = \frac{\omega_{*a}}{|\omega_{da,0}|}
\end{equation}
and
\begin{equation}
\sigma  = \frac{\omega_{da,0}}{|\omega_{da,0}|},
\end{equation}
we can recast \eq{eq:h_equation_local_linear_ITG_with_impurity_ions} into
\begin{eqnarray}\label{eq:h_equation_local_linear_ITG_with_impurity_ions_approx_and_new_coor}
\fl
\check h_a
=
\frac{
\Omega_a -  \Omega_{*a} \left[
1 + \eta_a
\left(
v_{||}^2 + v_\perp^2 - 3/2
\right)
\right]
}
{
\Omega_a - v_{||} K_{||} - \sigma (v_{||}^2 + v_\perp^2/2)
}
\frac{Z_a e  \check \varphi}{T_a}
J_0\left(\sqrt{2b_a} \, v_\perp\right)
f_{Ma},
\end{eqnarray}
where
\begin{equation}
b_a = \frac{k_\perp^2 m_a T_a}{Z_a^2 e^2 B^2}
\end{equation}
and
\begin{equation}
f_{Ma} = n_a\left(
\frac{m_a}{2\pi T_a}
\right)^{3/2}\exp\left(-v_{||}^2 - v_\perp^2\right).
\end{equation}

Plugging \eq{eq:h_equation_local_linear_ITG_with_impurity_ions_approx_and_new_coor} into the quasineutrality equation \eq{eq:quasineutrality_h_linear}, we find the dispersion relation
\begin{eqnarray}\label{eq:dispersion_relation}
\frac{Z_i^2 n_i}{T_i}
\frac{T_e}{n_e}
+
\frac{Z_z^2 n_z}{T_z}
\frac{T_e}{n_e}
+
1
 -
 \sum_a 
 D_a
 =
0,
\end{eqnarray}
with $n_e = Z_i n_i + Z_z n_z$ and
\begin{equation}\label{eq:def_Da}
D_a :=  \frac{\pi v_{ta}^3Z_aT_e}{e n_e \check \varphi}
 \int
 \check h_a 
J_0\left(\sqrt{2b_a} \, v_\perp\right)
\dd v_{||}\dd v_\perp^2
.
\end{equation}

The quantity $D_a$ has been computed in \cite{Parisi2020}. The result is
\begin{eqnarray}\label{eq:result_Da}
\fl
D_a
=
i Z_a^2\frac{T_e n_a}{T_a n_e}
\int_0^\infty
\dd\lambda
\frac{\Gamma_0({\hat b}^\sigma_a)}{(1+i\sigma\lambda)^{1/2}}\frac{1}{1+i\sigma\lambda/2} \exp\left(i\lambda\Omega_a - \frac{(\lambda K_{||})^2}{4(1+ i \sigma \lambda)}
\right)
\\[5pt]
\fl
\hspace{0.1cm}
\times
\left\{
\Omega_{*a}\left[
1 + \eta_a \left(
\frac{1
+
{\hat b}^\sigma_a\left(\Gamma_1({\hat b}^\sigma_a) / \Gamma_0({\hat b}^\sigma_a) - 1\right)
}{1+i\sigma\lambda/2} + \frac{2(1+i\sigma\lambda) - (K_{||}\lambda)^2}{4(1+i\sigma\lambda)^{2}}
- \frac{3}{2}
\right)
\right]
-\Omega_a
\right\}.
\nonumber
\end{eqnarray}
Here,
\begin{equation}
\Gamma_\nu(x) = I_\nu(x) \exp(-x),
\end{equation}
where $I_\nu$ denotes the modified Bessel function of the first kind of order $\nu$ and
\begin{equation}
{\hat b}^\sigma_a = \frac{b_a}{1 + i\sigma\lambda/2}.
\end{equation}
In \cite{Parisi2020}, a deformation of the integration path of the integral in \eq{eq:result_Da} is proposed that makes the numerical calculation of $D_a$ more efficient.

\section{Benchmark of the \texttt{python} script that solves the local dispersion relation}
\label{sec:benchmark_python_script}

\begin{figure}[h!]
	\centering  
	\includegraphics[width=\textwidth]{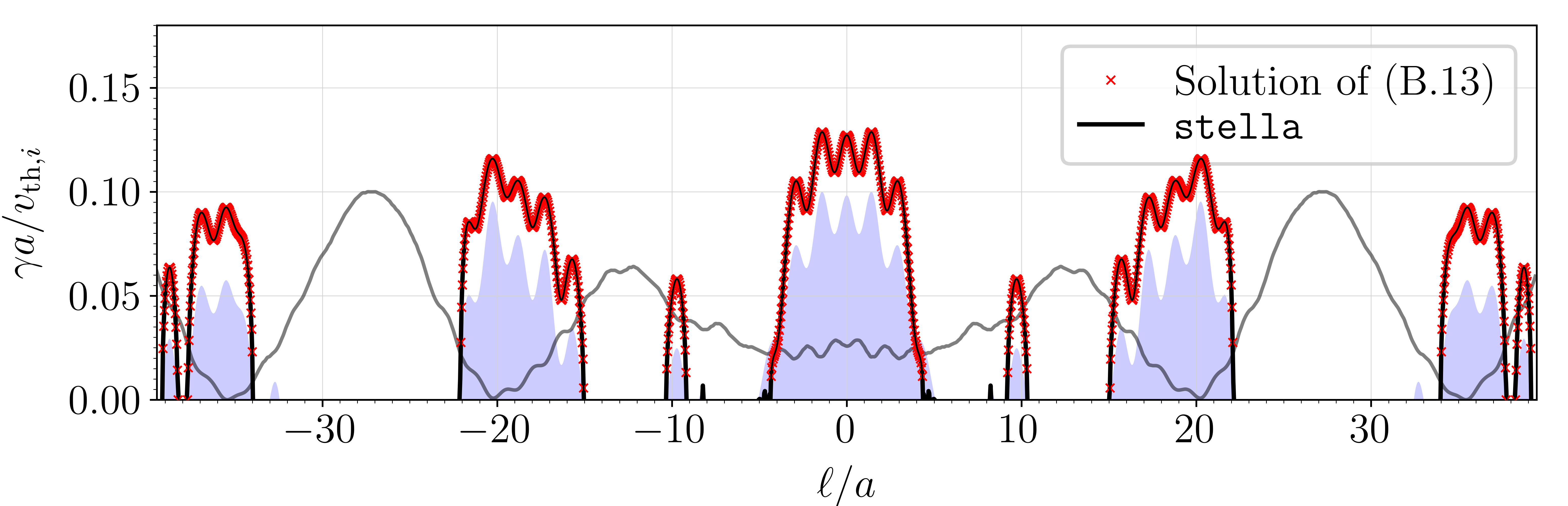} 
	\vspace{-8mm}
	\caption{Comparison, in W7-X geometry, between the growth rate given by the solution to \eq{eq:dispersion_relation} with $k_{||} = 0$ (red points) and the growth rate obtained from linear \texttt{stella} simulations without parallel streaming terms (thick black curve). We consider hydrogen bulk ions with $a/L_{T_i} = 3$, $a/L_{n_i} = 1$, and we take $T_i = T_e$ . No impurities are included. The range shown in $\ell/a$ corresponds to one  poloidal  turn and, in this case, $N_\ell = 2024$ has been taken. For reference, the structure of the magnetic field strength along the flux tube and the location of bad curvature regions have been included, following the conventions of figure \ref{fig:geometry_W7-X_LHD}.}
	\label{fig:benchmark_python_script}
\end{figure}

Here, we check that the \texttt{python} script written to solve the local dispersion relation \eq{eq:dispersion_relation} (with $D_a$ given by \eq{eq:result_Da}) gives correct results. This is not intended to be an exhaustive benchmark of the script. In figure \ref{fig:benchmark_python_script}, we perform a simple comparison between the local growth rate given by the solution to \eq{eq:dispersion_relation} for $k_{||} = 0$ and the growth rate obtained from linear \texttt{stella} simulations without parallel streaming terms. The agreement is excellent.

\section{Accuracy of the model local in $\ell$  including a finite parallel wavenumber}
\label{sec:accuracy_model_finite_k_par}

The results shown in \ref{sec:accuracy_toroidal_ITG_equations}, obtained using the local-in-$\ell$ toroidal ITG dispersion relation, can be improved by replacing equation \eq{eq:h_equation_local_linear_toroidal} by \eq{eq:h_equation_local_linear_finite_k_par}, allowing a finite $k_{||}$.

\begin{figure}[h!]
\centering
\hspace*{-4.5mm}
\includegraphics{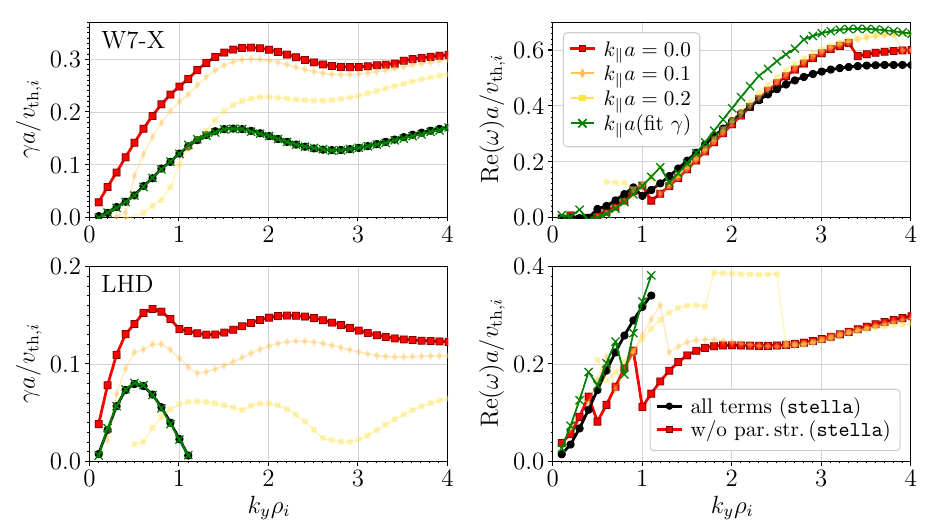}
\caption{Growth rate and real frequency in W7-X and LHD as a function of $k_y$. The red curves correspond to the values predicted by the local-in-$\ell$ dispersion relation for the toroidal branch of the ITG mode and the black curves correspond to the exact values. The effect of adding a finite $k_\parallel$ (the same for all values of $k_y$) to the local-in-$\ell$ toroidal ITG dispersion relation is illustrated by the faint orange and yellow lines. The green curves are obtained by choosing, for each $k_y$, the value of $k_{\parallel}$  that matches the exact growth rate value.}
\label{fig:gamma_omega_finite_and_const_kpar}
\end{figure}

\begin{figure}[h!]
\centering
\includegraphics{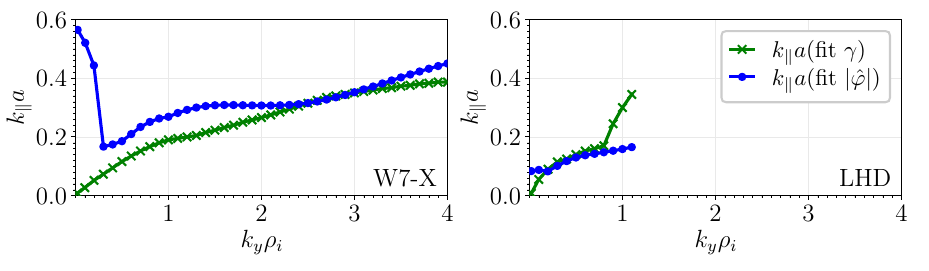}
\vspace{-9mm}
\caption{In green, the parallel wavenumber used to get the green curves in figure \ref{fig:gamma_omega_finite_and_const_kpar}. In blue, the parallel wavenumber obtained from fitting the parallel mode structure in complete linear \texttt{stella} simulations to a function of the form $\cos(k_{||} \ell)$.}
\label{fig:gamma_omega_finite_and_const_AND_correct_kpar}
\end{figure}

Let us go back to the cases discussed in \ref{sec:accuracy_toroidal_ITG_equations}, specifically in figure \ref{fig:examples_accuracy_toroidal_ITG_equations}, and add a finite $k_{||}$. In figure \ref{fig:gamma_omega_finite_and_const_kpar}, orange and yellow lines correspond to  adding the same value of $k_{||}$ for all values of $k_y$. This illustrates the qualitative effect of $k_{||}$, which is small on the frequency and tends to reduce $\gamma$. Of course, a meaningful choice for $k_{||}$ should depend on $k_y$. Green curves correspond to choosing $k_{||}$ at each $k_y$ so that the growth rate $\gamma$ obtained from the local dispersion relation exactly matches $\gamma$ computed from complete linear simulations with \texttt{stella}. We see that the frequency predicted by the local-in-$\ell$ dispersion relation is close to the one obtained from \texttt{stella} even with this finite $k_{||}$. In figure \ref{fig:gamma_omega_finite_and_const_AND_correct_kpar}, we check the consistency of the model. The choice for $k_{||}$ just explained is compared with the value of $k_{||}$ that gives the best fit (in the region where the mode peaks) of the parallel mode structure of the exact ITG mode in a complete linear simulation with \texttt{stella} to a function of the form $\cos(k_{||} \ell)$. Consistency holds in broad ranges of $k_y$ but clearly breaks at sufficiently small values of $k_y$, which is expected because, for very small $k_y$, modes tend to delocalize~\cite{Plunk2014, Zocco2018}.

\section{Dispersion relation of the toroidal ITG mode for small Larmor radius}
\label{sec:dispersion_relation_small_krho}

The calculation presented below is similar to that carried out in \cite{Biglari1989}. In \eq{eq:h_equation_local_linear_ITG_with_impurity_ions}, we assume $k_{||} = 0$ and $k_\perp^2 \rho_a^2 \ll 1$ so that $J_0(k_\perp \rho_a)\simeq 1$.  From now on, and for definiteness, we assume that $\omega_{da, 0} > 0$. The sign of $\omega_{da, 0}$ can be chosen by selecting the sign of $\bk_\perp$, and this does not reduce the generality of our calculation because reality of the solution of the gyrokinetic equations implies $\omega_{-\bk_\perp} = - \omega*_{\bk_\perp}$, where the asterisk stands for complex conjugation.
Then, \eq{eq:h_equation_local_linear_ITG_with_impurity_ions} becomes
\begin{eqnarray}\label{eq:h_equation_local_linear_ITG_with_impurity_ions_approx_and_new_coor_2}
\fl
\check h_a
=
n_a\left(
\frac{m_a}{2\pi T_a}
\right)^{3/2}
\frac{Z_a e  \check \varphi^{\rm tb}}{T_a}
\,
\frac{
\Omega_a -  \Omega_{*a} \left[
1 + \eta_a
\left(
v_{||}^2 + U_\perp - 3/2
\right)
\right]
}
{
\Omega_a - v_{||}^2 - U_\perp/2
}
\exp\left(-v_{||}^2 - U_\perp\right),
\end{eqnarray}
where we have written $U_\perp = v_\perp^2$ and we have employed notation introduced in \ref{sec:local_dispersion_relation}. Plugging \eq{eq:h_equation_local_linear_ITG_with_impurity_ions_approx_and_new_coor_2} into the quasineutrality equation \eq{eq:quasineutrality_h_linear} and using again that $J_0(k_\perp \rho_a)\simeq 1$, we arrive at the dispersion relation
\begin{eqnarray}\label{eq:quasineutrality_h_linear_ITG_with_impurity_ions_new_coor}
\frac{Z_i^2 n_i}{T_i}
\frac{T_e}{n_e}
+
\frac{Z_z^2 n_z}{T_z}
\frac{T_e}{n_e}
+
1
-
\sum_a
\frac{T_e}{n_e}
 \frac{Z_a^2 n_a}{T_a}
I_a
=
0,
\end{eqnarray}
where
\begin{eqnarray}\label{eq:integral_I}
\fl
I_a
=
\frac{1}{\sqrt{\pi}}
\int_{-\infty}^\infty\dd v_{||}\int_0^\infty\dd U_\perp \,
\frac{
\Omega_a -  \Omega_{*a} \left[
1 + \eta_a
\left(
v_{||}^2 + U_\perp - 3/2
\right)
\right]
}
{
\Omega_a - v_{||}^2 - U_\perp/2
}
\exp\left(-v_{||}^2 - U_\perp\right).
\end{eqnarray}

The right-hand side of \eq{eq:integral_I} can be written as
\begin{eqnarray}\label{eq:integral_I_rearranged}
I_a
=
2 \Omega_{*a}\eta_a
+
\left\{
\Omega_a - \Omega_{*a}\left[ 1+ \eta_a \left(2\Omega_a - 3/2\right)\right]
\right\}
F_a 
+ \Omega_{*a} \eta_a G_a
,
\end{eqnarray}
where
\begin{eqnarray}\label{eq:integral_F}
F_a
=
\frac{1}{\sqrt{\pi}}
\int_{-\infty}^\infty\dd v_{||}\int_0^\infty\dd U_\perp \,
\frac{
\exp(-v_{||}^2 - U_\perp)
}
{
\Omega_a - v_{||}^2 - U_\perp/2
}
\end{eqnarray}
and
\begin{eqnarray}\label{eq:integral_G}
G_a
=
\frac{1}{\sqrt{\pi}}
\int_{-\infty}^\infty\dd v_{||}\int_0^\infty\dd U_\perp \,
\frac{
v_{||}^2 \exp(-v_{||}^2 - U_\perp)
}
{
\Omega_a - v_{||}^2 - U_\perp/2
}
\,
.
\end{eqnarray}

The integrals $F_a$ and $G_a$ can be expressed in terms of the plasma dispersion function,
\begin{equation}\label{eq:definition_plasma_dispersion_function}
{\cal Z}(\zeta) = \frac{1}{\sqrt{\pi}} \int_{-\infty}^\infty \frac{\exp(-v_{||}^2)}{v_{||} - \zeta} \dd v_{||},
\end{equation}
where the path of integration is the one given by Landau~\cite{Landau1946}. First, we find two ordinary differential equations satisfied by $F_a$ and $G_a$. We start by writing $F_a$ as a function of $\zeta = \Omega_a^{1/2}$,
\begin{eqnarray}\label{eq:integral_F_in_terms_of_zeta}
F_a(\zeta)
=
\frac{1}{\sqrt{\pi}}
\int_{-\infty}^\infty\dd v_{||}\int_0^\infty\dd U_\perp \,
\frac{
\exp(-v_{||}^2 - U_\perp)
}
{
\zeta^2 - v_{||}^2 - U_\perp/2
}\,
.
\end{eqnarray}
\textcolor{black}{Here, $\Omega_a^{1/2}$ denotes the square root of $\Omega_a$ with argument in $[0,\pi)$.}

Differentiating with respect to $\zeta$,
\begin{eqnarray}\label{eq:dFdzeta}
\fl
\frac{\dd F_a}{\dd \zeta}
=
-\frac{2\zeta}{\sqrt{\pi}}
\int_{-\infty}^\infty\dd v_{||}\int_0^\infty\dd U_\perp \,
\frac{
\exp(-v_{||}^2 - U_\perp)
}
{
(\zeta^2 - v_{||}^2 - U_\perp/2)^2
}
=
\nonumber\\[5pt]
\hspace{1cm}
\fl
-\frac{4\zeta}{\sqrt{\pi}}
\int_{-\infty}^\infty\dd v_{||}\int_0^\infty\dd U_\perp \,
\partial_{U_\perp}
\left(
\frac{
1
}
{
\zeta^2 - v_{||}^2 - U_\perp/2
}
\right)
\exp(-v_{||}^2 - U_\perp)
=
\nonumber\\[5pt]
\hspace{1cm}
\fl
\frac{4\zeta}{\sqrt{\pi}}
\int_{-\infty}^\infty
\frac{
\exp(-v_{||}^2)
}
{
\zeta^2 - v_{||}^2
}
\dd v_{||}
-\frac{4\zeta}{\sqrt{\pi}}
\int_{-\infty}^\infty\dd v_{||}\int_0^\infty\dd U_\perp \,
\frac{
\exp(-v_{||}^2 - U_\perp)
}
{
\zeta^2 - v_{||}^2 - U_\perp/2
}\,
.
\end{eqnarray}
In the last step, we have integrated by parts in $U_\perp$. We have found that $F_a$ satisfies the ordinary differential equation
\begin{eqnarray}\label{eq:ODE_F_preliminary}
\frac{\dd F_a}{\dd \zeta}
=
-4\zeta F_a
+
\frac{4\zeta}{\sqrt{\pi}}
\int_{-\infty}^\infty
\frac{
\exp(-v_{||}^2)
}
{
\zeta^2 - v_{||}^2
}\dd v_{||}
.
\end{eqnarray}
Note that
\begin{eqnarray}\label{eq:dFdzeta_aux_term}
\fl
\frac{4\zeta}{\sqrt{\pi}}
\int_{-\infty}^\infty
\frac{
\exp(-v_{||}^2)
}
{
\zeta^2 - v_{||}^2
}
\dd v_{||}
=
\nonumber\\[5pt]
\hspace{1cm}
\fl
\frac{2}{\sqrt{\pi}}
\int_{-\infty}^\infty
\frac{
\exp(-v_{||}^2)
}
{
\zeta + v_{||}
}
\dd v_{||}
+
\frac{2}{\sqrt{\pi}}
\int_{-\infty}^\infty
\frac{
\exp(-v_{||}^2)
}
{
\zeta - v_{||}
}
\dd v_{||}
=
-4{\cal Z}(\zeta),
\end{eqnarray}
so that \eq{eq:ODE_F_preliminary} can be written as
\begin{eqnarray}\label{eq:ODE_F}
\frac{\dd F_a}{\dd \zeta}
+4\zeta F_a
=
-4{\cal Z}
.
\end{eqnarray}
Hence, the right-hand side of \eq{eq:integral_F_in_terms_of_zeta} satisfies \eq{eq:ODE_F}. Employing the relation
\begin{equation}
{\cal Z}'(\zeta) = -2\left(\zeta{\cal Z}(\zeta) +1\right),
\end{equation}
it is easy to show that ${\cal Z}^2(\zeta)$ also satisfies \eq{eq:ODE_F}. The integrals on the right-hand side of \eq{eq:integral_F_in_terms_of_zeta} for $\zeta = 0$ can be worked out analytically, obtaining $F_a(0) =  - \pi$. Noting that ${\cal Z}^2(0) = - \pi$, and due to the uniqueness of the solution of \eq{eq:ODE_F}, we deduce that the right-hand side of \eq{eq:integral_F_in_terms_of_zeta} equals ${\cal Z}^2(\zeta)$ and, finally, going back to \eq{eq:integral_F}, we conclude that
\begin{equation}\label{eq:F_result}
F_a = {\cal Z}^2(\Omega_a^{1/2}).
\end{equation}

Writing $G_a$ as a function of $\zeta = \Omega_a^{1/2}$ and after a computation analogous to \eq{eq:dFdzeta}, we get the ordinary differential equation for $G_a(\zeta)$
\begin{eqnarray}\label{eq:ODE_G}
\frac{\dd G_a}{\dd \zeta}
+4\zeta G_a
=
2\zeta{\cal Z}'.
\end{eqnarray}
From here, using $G_a(0) = \pi/2 - 2$, we infer that
\begin{equation}\label{eq:G_result}
G_a = -2 -2 \Omega_a^{1/2}{\cal Z}(\Omega_a^{1/2}) -\frac{1}{2}{\cal Z}^2(\Omega_a^{1/2}).
\end{equation}

Inserting \eq{eq:F_result} and \eq{eq:G_result} into \eq{eq:integral_I_rearranged}, we arrive at the result for $I_a$,
\begin{eqnarray}\label{eq:integral_I_final}
\fl
\hspace{1.5cm}
I_a
=
\left\{
\Omega_a - \Omega_{*a}\left[ 1+ \eta_a \left(2\Omega_a - 1\right)\right]
\right\}
{\cal Z}^2(\Omega_a^{1/2})
-2 \Omega_{*a} \eta_a \Omega_a^{1/2}{\cal Z}(\Omega_a^{1/2}).
\end{eqnarray}

Finally, we point out that the analytical calculations of section \ref{sec:solution_ITG_dispersion_relation} rely on the expansion of $I_a$ in certain limits. For this, the expansion of the plasma dispersion function \eq{eq:definition_plasma_dispersion_function} for large values of its argument is required. If $|\zeta| \gg 1$ and ${\rm Im}(\zeta) > 0$, the expansion reads~\cite{Fried_book}
\begin{equation}\label{eq:expansion_Z_large_zeta_Imzeta_positive}
{\cal Z}(\zeta) = -\frac{1}{\zeta}\left(
1 + \frac{1}{2\zeta^2} + \frac{3}{4\zeta^4} + \dots
\right).
\end{equation}

\section*{References}

\begin{thebibliography}{10}
\expandafter\ifx\csname url\endcsname\relax
  \def\url#1{{\tt #1}}\fi
\expandafter\ifx\csname urlprefix\endcsname\relax\def\urlprefix{URL }\fi
\providecommand{\eprint}[2][]{\url{#2}}

\bibitem{Helander2012}
Helander P, Beidler C~D, Bird T~M, Drevlak M, Feng Y, Hatzky R, Jenko F,
  Kleiber R, Proll J~H~E, Turkin Y and Xanthopoulos P 2012 {\em Plasma Physics
  and Controlled Fusion\/} {\bf 54} 124009

\bibitem{Calvo_2017}
Calvo I, Parra F~I, Velasco J~L and Alonso J~A 2017 {\em Plasma Physics and
  Controlled Fusion\/} {\bf 59} 055014

\bibitem{Herbemont_2022}
d'Herbemont V, Parra F~I, Calvo I and Velasco J~L 2022 {\em Journal of Plasma
  Physics\/} {\bf 88} 905880507

\bibitem{Dinklage_nf_2013}
Dinklage A, Yokoyama M, Tanaka K, Velasco J~L, L\'opez-Bruna D, Beidler C~D,
  Satake S, Ascas\'ibar E, Ar\'evalo J, Baldzuhn J, Feng Y, Gates D, Geiger J,
  Ida K, Jakubowski M, L\'opez-Fraguas A, Maassberg H, Miyazawa J, Morisaki T,
  Murakami S, Pablant N, Kobayashi S, Seki R, Suzuki C, Suzuki Y, Turkin Y,
  Wakasa A, Wolf R, Yamada H, Yoshinuma M, {LHD Exp Group}, {TJ-II Team} and
  {W7-AS Team} 2013 {\em Nuclear Fusion\/} {\bf 53} 063022

\bibitem{Bosch2017}
Bosch H~S, Brakel R, Braeuer T, Bykov V, van Eeten P, Feist J~H, F\"{u}llenbach
  F, Gasparotto M, Grote H, Klinger T, Laqua H, Nagel M, Naujoks D, Otte M,
  Risse K, Rummel T, Schacht J, Spring A, Sunn~Pedersen T, Vilbrandt R, Wegener
  L, Werner A, Wolf R, Baldzuhn J, Biedermann C, Braune H, Burhenn R, Hirsch M,
  H\"{o}fel U, Knauer J, Kornejew P, Marsen S, Stange T and Trimino~Mora H 2017
  {\em Nuclear Fusion\/} {\bf 57} 116015

\bibitem{Wolf_nf_2017}
Wolf R, Ali A, Alonso A, Baldzuhn J, Beidler C, Beurskens M, Biedermann C,
  Bosch H~S, Bozhenkov S, Brakel R, Dinklage A, Feng Y, Fuchert G, Geiger J,
  Grulke O {\em et~al.\/} 2017 {\em Nuclear Fusion\/} {\bf 57} 102020

\bibitem{Klinger2019}
Klinger T, Andreeva T, Bozhenkov S, Brandt C, Burhenn R, Buttensch\"{o}n B,
  Fuchert G, Geiger B, Grulke O, Laqua H, Pablant N, Rahbarnia K, Stange T, von
  Stechow A, Tamura N, Thomsen H, Turkin Y, Wegner T, Abramovic I,
  \"{A}k\"{a}slompolo S, Alcuson J, Aleynikov P, Aleynikova K, Ali A, Alonso A,
  Anda G, Ascasibar E, B\"{a}hner J, Baek S, Balden M, Baldzuhn J, Banduch M,
  Barbui T, Behr W, Beidler C, Benndorf A, Biedermann C, Biel W, Blackwell B,
  Blanco E, Blatzheim M, Ballinger S, Bluhm T, B\"{o}ckenhoff D, B\"{o}swirth
  B, B\"{o}ttger L~G, Borchardt M, Borsuk V, Boscary J, Bosch H~S, Beurskens M,
  Brakel R, Brand H, Br\"{a}uer T, Braune H, Brezinsek S, Brunner K~J, Bussiahn
  R, Bykov V, Cai J, Calvo I, Cannas B, Cappa A, Carls A, Carralero D, Carraro
  L, Carvalho B, Castejon F, Charl A, Chaudhary N, Chauvin D, Chernyshev F,
  Cianciosa M, Citarella R, Claps G, Coenen J, Cole M, Cole M, Cordella F, Cseh
  G, Czarnecka A, Czerski K, Czerwinski M, Czymek G, da~Molin A, da~Silva A,
  Damm H, de~la Pena A, Degenkolbe S, Dhard C, Dibon M, Dinklage A, Dittmar T,
  Drevlak M, Drewelow P, Drews P, Durodie F, Edlund E, van Eeten P, Effenberg
  F, Ehrke G, Elgeti S, Endler M, Ennis D, Esteban H, Estrada T, Fellinger J,
  Feng Y, Flom E, Fernandes H, Fietz W, Figacz W, Fontdecaba J, Ford O, Fornal
  T, Frerichs H, Freund A, Funaba T, Galkowski A, Gantenbein G, Gao Y,
  García~Regaña J, Gates D, Geiger J, Giannella V, Gogoleva A, Goncalves B,
  Goriaev A, Gradic D, Grahl M, Green J, Greuner H, Grosman A, Grote H, Gruca
  M, Guerard C, Hacker P, Han X, Harris J, Hartmann D, Hathiramani D, Hein B,
  Heinemann B, Helander P, Henneberg S, Henkel M, Hernandez~Sanchez J, Hidalgo
  C, Hirsch M, Hollfeld K, H\"{o}fel U, H\"{o}lting A, H\"{o}schen D, Houry M,
  Howard J, Huang X, Huang Z, Hubeny M, Huber M, Hunger H, Ida K, Ilkei T, Illy
  S, Israeli B, Jablonski S, Jakubowski M, Jelonnek J, Jenzsch H, Jesche T, Jia
  M, Junghanns P, Kacmarczyk J, Kallmeyer J~P, Kamionka U, Kasahara H, Kasparek
  W, Kazakov Y, Kenmochi N, Killer C, Kirschner A, Kleiber R, Knauer J, Knaup
  M, Knieps A, Kobarg T, Kocsis G, K\"{o}chl F, Kolesnichenko Y, K\"{o}nies A,
  K\"{o}nig R, Kornejew P, Koschinsky J~P, K\"{o}ster F, Kr\"{a}mer M, Krampitz
  R, Kr\"{a}mer-Flecken A, Krawczyk N, Kremeyer T, Krom J, Krychowiak M,
  Ksiazek I, Kubkowska M, K\"{u}hner G, Kurki-Suonio T, Kurz P, Kwak S,
  Landreman M, Lang P, Lang R, Langenberg A, Langish S, Laqua H, Laube R,
  Lazerson S, Lechte C, Lennartz M, Leonhardt W, Li C, Li C, Li Y, Liang Y,
  Linsmeier C, Liu S, Lobsien J~F, Loesser D, Loizu~Cisquella J, Lore J, Lorenz
  A, Losert M, L\"{u}cke A, Lumsdaine A, Lutsenko V, Maassberg H, Marchuk O,
  Matthew J, Marsen S, Marushchenko M, Masuzaki S, Maurer D, Mayer M, McCarthy
  K, McNeely P, Meier A, Mellein D, Mendelevitch B, Mertens P, Mikkelsen D,
  Mishchenko A, Missal B, Mittelstaedt J, Mizuuchi T, Mollen A, Moncada V,
  M\"{o}nnich T, Morisaki T, Moseev D, Murakami S, Náfrádi G, Nagel M, Naujoks
  D, Neilson H, Neu R, Neubauer O, Neuner U, Ngo T, Nicolai D, Nielsen S,
  Niemann H, Nishizawa T, Nocentini R, N\"{u}hrenberg C, N\"{u}hrenberg J,
  Obermayer S, Offermanns G, Ogawa K, \"{O}lmanns J, Ongena J, Oosterbeek J,
  Orozco G, Otte M, Pacios~Rodriguez L, Panadero N, Panadero~Alvarez N,
  Papenfuß D, Paqay S, Pasch E, Pavone A, Pawelec E, Pedersen T, Pelka G,
  Perseo V, Peterson B, Pilopp D, Pingel S, Pisano F, Plaum B, Plunk G,
  P\"{o}l\"{o}skei P, Porkolab M, Proll J, Puiatti M~E, Puig~Sitjes A, Purps F,
  Rack M, R\'ecsei S, Reiman A, Reimold F, Reiter D, Remppel F, Renard S, Riedl
  R, Riemann J, Risse K, Rohde V, R\"{o}hlinger H, Rom\'e M, Rondeshagen D, Rong
  P, Roth B, Rudischhauser L, Rummel K, Rummel T, Runov A, Rust N, Ryc L,
  Ryosuke S, Sakamoto R, Salewski M, Samartsev A, Sanchez E, Sano F, Satake S,
  Schacht J, Satheeswaran G, Schauer F, Scherer T, Schilling J, Schlaich A,
  Schlisio G, Schluck F, Schl\"{u}ter K~H, Schmitt J, Schmitz H, Schmitz O,
  Schmuck S, Schneider M, Schneider W, Scholz P, Schrittwieser R, Schr\"{o}der
  M, Schr\"{o}der T, Schroeder R, Schumacher H, Schweer B, Scott E, Sereda S,
  Shanahan B, Sibilia M, Sinha P, Sipli\"{a} S, Slaby C, Sleczka M, Smith H,
  Spiess W, Spong D, Spring A, Stadler R, Stejner M, Stephey L, Stridde U,
  Suzuki C, Svensson J, Szab\'o V, Szabolics T, Szepesi T, Sz\"{o}kefalvi-Nagy Z,
  Tancetti A, Terry J, Thomas J, Thumm M, Travere J, Traverso P, Tretter J,
  Trimino~Mora H, Tsuchiya H, Tsujimura T, Tulip\'an S, Unterberg B, Vakulchyk I,
  Valet S, Vano L, van Milligen B, van Vuuren A, Vela L, Velasco J~L, Vergote
  M, Vervier M, Vianello N, Viebke H, Vilbrandt R, Vork\"{o}per A, Wadle S,
  Wagner F, Wang E, Wang N, Wang Z, Warmer F, Wauters T, Wegener L, Weggen J,
  Wei Y, Weir G, Wendorf J, Wenzel U, Werner A, White A, Wiegel B, Wilde F,
  Windisch T, Winkler M, Winter A, Winters V, Wolf S, Wolf R, Wright A, Wurden
  G, Xanthopoulos P, Yamada H, Yamada I, Yasuhara R, Yokoyama M, Zanini M,
  Zarnstorff M, Zeitler A, Zhang D, Zhang H, Zhu J, Zilker M, Zocco A, Zoletnik
  S and Zuin M 2019 {\em Nuclear Fusion\/} {\bf 59} 112004

\bibitem{Bozhenkov2020}
Bozhenkov S~A, Kazakov Y, Ford O~P, Beurskens M~N~A, Alcus\'on J, Alonso J~A,
  Baldzuhn J, Brandt C, Brunner K~J, Damm H, Fuchert G, Geiger J, Grulke O,
  Hirsch M, H\"{o}fel U, Huang Z, Knauer J, Krychowiak M, Langenberg A, Laqua
  H~P, Lazerson S, B~Marushchenko N, Moseev D, Otte M, Pablant N, Pasch E,
  Pavone A, Proll J~H~E, Rahbarnia K, Scott E~R, Smith H~M, Stange T, von
  Stechow A, Thomsen H, Turkin Y, Wurden G, Xanthopoulos P, Zhang D and Wolf
  R~C 2020 {\em Nuclear Fusion\/} {\bf 60} 066011

\bibitem{Beurskens_2021}
Beurskens M, Bozhenkov S, Ford O, Xanthopoulos P, Zocco A, Turkin Y, Alonso A,
  Beidler C, Calvo I, Carralero D, Estrada T, Fuchert G, Grulke O, Hirsch M,
  Ida K, Jakubowski M, Killer C, Krychowiak M, Kwak S, Lazerson S, Langenberg
  A, Lunsford R, Pablant N, Pasch E, Pavone A, Reimold F, Romba T, von Stechow
  A, Smith H, Windisch T, Yoshinuma M, Zhang D, Wolf R and {the W7-X Team} 2021
  {\em Nuclear Fusion\/} {\bf 61} 116072

\bibitem{Beidler2021}
Beidler C~D, Smith H~M, Alonso A, Andreeva T, Baldzuhn J, Beurskens M~N~A,
  Borchardt M, Bozhenkov S~A, Brunner K~J, Damm H, Drevlak M, Ford O~P, Fuchert
  G, Geiger J, Helander P, Hergenhahn U, Hirsch M, H\"{o}fel U, Kazakov Y~O,
  Kleiber R, Krychowiak M, Kwak S, Langenberg A, Laqua H~P, Neuner U, Pablant
  N~A, Pasch E, Pavone A, Pedersen T~S, Rahbarnia K, Schilling J, Scott E~R,
  Stange T, Svensson J, Thomsen H, Turkin Y, Warmer F, Wolf R~C, Zhang D,
  Abramovic I, \"{A}k\"{a}slompolo S, Alcus\'on J, Aleynikov P, Aleynikova K, Ali
  A, Alonso A, Anda G, Ascasibar E, B\"{a}hner J~P, Baek S~G, Balden M, Banduch
  M, Barbui T, Behr W, Benndorf A, Biedermann C, Biel W, Blackwell B, Blanco E,
  Blatzheim M, Ballinger S, Bluhm T, B\"{o}ckenhoff D, B\"{o}swirth B,
  B\"{o}ttger L~G, Borsuk V, Boscary J, Bosch H~S, Brakel R, Brand H, Brandt C,
  Br\"{a}uer T, Braune H, Brezinsek S, Brunner K~J, Burhenn R, Bussiahn R,
  Buttensch\"{o}n B, Bykov V, Cai J, Calvo I, Cannas B, Cappa A, Carls A,
  Carraro L, Carvalho B, Castejon F, Charl A, Chaudhary N, Chauvin D,
  Chernyshev F, Cianciosa M, Citarella R, Claps G, Coenen J, Cole M, Cole M~J,
  Cordella F, Cseh G, Czarnecka A, Czerski K, Czerwinski M, Czymek G, da~Molin
  A, da~Silva A, de~la Pena A, Degenkolbe S, Dhard C~P, Dibon M, Dinklage A,
  Dittmar T, Drewelow P, Drews P, Durodie F, Edlund E, Effenberg F, Ehrke G,
  Elgeti S, Endler M, Ennis D, Esteban H, Estrada T, Fellinger J, Feng Y, Flom
  E, Fernandes H, Fietz W~H, Figacz W, Fontdecaba J, Fornal T, Frerichs H,
  Freund A, Funaba T, Galkowski A, Gantenbein G, Gao Y, Garc\'{\i}a~Rega\~na J, Gates
  D, Geiger B, Giannella V, Gogoleva A, Goncalves B, Goriaev A, Gradic D, Grahl
  M, Green J, Greuner H, Grosman A, Grote H, Gruca M, Grulke O, Guerard C,
  Hacker P, Han X, Harris J~H, Hartmann D, Hathiramani D, Hein B, Heinemann B,
  Helander P, Henneberg S, Henkel M, Hergenhahn U, Hernandez~Sanchez J, Hidalgo
  C, Hollfeld K~P, H\"{o}lting A, H\"{o}schen D, Houry M, Howard J, Huang X,
  Huang Z, Hubeny M, Huber M, Hunger H, Ida K, Ilkei T, Illy S, Israeli B,
  Jablonski S, Jakubowski M, Jelonnek J, Jenzsch H, Jesche T, Jia M, Junghanns
  P, Kacmarczyk J, Kallmeyer J~P, Kamionka U, Kasahara H, Kasparek W, Kenmochi
  N, Killer C, Kirschner A, Klinger T, Knauer J, Knaup M, Knieps A, Kobarg T,
  Kocsis G, K\"{o}chl F, Kolesnichenko Y, K\"{o}nies A, K\"{o}nig R, Kornejew
  P, Koschinsky J~P, K\"{o}ster F, Kr\"{a}mer M, Krampitz R, Kr\"{a}mer-Flecken
  A, Krawczyk N, Kremeyer T, Krom J, Ksiazek I, Kubkowska M, K\"{u}hner G,
  Kurki-Suonio T, Kurz P~A, Landreman M, Lang P, Lang R, Langish S, Laqua H,
  Laube R, Lazerson S, Lechte C, Lennartz M, Leonhardt W, Li C, Li C, Li Y,
  Liang Y, Linsmeier C, Liu S, Lobsien J~F, Loesser D, Loizu~Cisquella J, Lore
  J, Lorenz A, Losert M, L\"{u}cke A, Lumsdaine A, Lutsenko V, Maassberg H,
  Marchuk O, Matthew J~H, Marsen S, Marushchenko M, Masuzaki S, Maurer D, Mayer
  M, McCarthy K, McNeely P, Meier A, Mellein D, Mendelevitch B, Mertens P,
  Mikkelsen D, Mishchenko A, Missal B, Mittelstaedt J, Mizuuchi T, Mollen A,
  Moncada V, M\"{o}nnich T, Morisaki T, Moseev D, Murakami S, N\'afr\'adi G, Nagel
  M, Naujoks D, Neilson H, Neu R, Neubauer O, Ngo T, Nicolai D, Nielsen S~K,
  Niemann H, Nishizawa T, Nocentini R, N\"{u}hrenberg C, N\"{u}hrenberg J,
  Obermayer S, Offermanns G, Ogawa K, \"{O}lmanns J, Ongena J, Oosterbeek J~W,
  Orozco G, Otte M, Pacios~Rodriguez L, Panadero N, Panadero~Alvarez N,
  Papenfuss D, Paqay S, Pawelec E, Pedersen T~S, Pelka G, Perseo V, Peterson B,
  Pilopp D, Pingel S, Pisano F, Plaum B, Plunk G, P\"{o}l\"{o}skei P, Porkolab
  M, Proll J, Puiatti M~E, Puig~Sitjes A, Purps F, Rack M, R\'ecsei S, Reiman A,
  Reimold F, Reiter D, Remppel F, Renard S, Riedl R, Riemann J, Risse K, Rohde
  V, R\"{o}hlinger H, Rom\'e M, Rondeshagen D, Rong P, Roth B, Rudischhauser L,
  Rummel K, Rummel T, Runov A, Rust N, Ryc L, Ryosuke S, Sakamoto R, Salewski
  M, Samartsev A, S\'anchez E, Sano F, Satake S, Schacht J, Satheeswaran G,
  Schauer F, Scherer T, Schlaich A, Schlisio G, Schluck F, Schl\"{u}ter K~H,
  Schmitt J, Schmitz H, Schmitz O, Schmuck S, Schneider M, Schneider W, Scholz
  P, Schrittwieser R, Schr\"{o}der M, Schr\"{o}der T, Schroeder R, Schumacher
  H, Schweer B, Sereda S, Shanahan B, Sibilia M, Sinha P, Sipli\"{a} S, Slaby
  C, Sleczka M, Spiess W, Spong D~A, Spring A, Stadler R, Stejner M, Stephey L,
  Stridde U, Suzuki C, Szab\'o V, Szabolics T, Szepesi T, Sz\"{o}kefalvi-Nagy Z,
  Tamura N, Tancetti A, Terry J, Thomas J, Thumm M, Travere J~M, Traverso P,
  Tretter J, Trimino~Mora H, Tsuchiya H, Tsujimura T, Tulip\'an S, Unterberg B,
  Vakulchyk I, Valet S, Van\'o L, van Eeten P, van Milligen B, van Vuuren A~J,
  Vela L, Velasco J~L, Vergote M, Vervier M, Vianello N, Viebke H, Vilbrandt R,
  von Stechow A, Vork\"{o}per A, Wadle S, Wagner F, Wang E, Wang N, Wang Z,
  Wauters T, Wegener L, Weggen J, Wegner T, Wei Y, Weir G, Wendorf J, Wenzel U,
  Werner A, White A, Wiegel B, Wilde F, Windisch T, Winkler M, Winter A,
  Winters V, Wolf S, Wolf R~C, Wright A, Wurden G, Xanthopoulos P, Yamada H,
  Yamada I, Yasuhara R, Yokoyama M, Zanini M, Zarnstorff M, Zeitler A, Zhang H,
  Zhu J, Zilker M, Zocco A, Zoletnik S and Zuin M 2021 {\em Nature\/} {\bf 596}
  221

\bibitem{McKee2000}
McKee G, Burrell K, Fonck R, Jackson G, Murakami M, Staebler G, Thomas D and
  West P 2000 {\em Physical Review Letters\/} {\bf 84} 1922

\bibitem{Osakabe2014}
Osakabe M, Takahashi H, Nagaoka K, Murakami S, Yamada I, Yoshinuma M, Ida K,
  Yokoyama M, Seki R, Lee H, Nakamura Y, Tamura N, Sudo S, Tanaka K, Seki T,
  Takeiri Y, Kaneko O and Yamada H 2014 {\em Plasma Physics and Controlled
  Fusion\/} {\bf 56} 095011

\bibitem{Takahashi2018}
Takahashi H, Nagaoka K, Mukai K, Yokoyama M, Murakami S, Ohdachi S, Bando T,
  Narushima Y, Nakano H, Osakabe M, Ida K, Yoshinuma M, Seki R, Yamaguchi H,
  Tanaka K, Nakata M, Warmer F, Oishi T, Goto M, Morita S, Tsujimura T, Kubo S,
  Kobayashi T, Yamada I, Suzuki C, Emoto M, Ido T, Shimizu A, Tokuzawa T,
  Nagasaki K, Morisaki T and and Y~T 2018 {\em Nuclear Fusion\/} {\bf 58}
  106028

\bibitem{Lunsford2021}
Lunsford R, Killer C, Nagy A, Gates D~A, Klinger T, Dinklage A, Satheeswaran G,
  Kocsis G, Lazerson S~A, Nespoli F, Pablant N~A, von Stechow A, Alonso A,
  Andreeva T, Beurskens M, Biedermann C, Brezinsek S, Brunner K~J,
  Buttensch\"{o}n B, Carralero D, Cseh G, Drewelow P, Effenberg F, Estrada T,
  Ford O~P, Grulke O, Hergenhahn U, H\"{o}fel U, Knauer J, Krause M, Krychowiak
  M, Kwak S, Langenberg A, Neuner U, Nicolai D, Pavone A, Puig~Sitjes A,
  Rahbarnia K, Schilling J, Svensson J, Szepesi T, Thomsen H, Wauters T,
  Windisch T, Winters V, Zhang D and Zsuga L 2021 {\em Physics of Plasmas\/}
  {\bf 28} 082506

\bibitem{Nespoli2022}
Nespoli F, Masuzaki S, Tanaka K, Ashikawa N, Shoji M, Gilson E~P, Lunsford R,
  Oishi T, Ida K, Yoshinuma M, Takemura Y, Kinoshita T, Motojima G, Kenmochi N,
  Kawamura G, Suzuki C, Nagy A, Bortolon A, Pablant N~A, Mollen A, Tamura N,
  Gates D~A and Morisaki T 2022 {\em Nature Physics\/} {\bf 18} 350

\bibitem{GarciaRegana2024}
Garc{\'{\i}}a-Rega\~na J~M, Calvo I, Parra F~I and Thienpondt H 2024 {\em
  Physical Review Letters\/} {\bf 133} 105101

\bibitem{Tang1980}
Tang W~M, White R~B and Guzdar P~N 1980 {\em The Physics of Fluids\/} {\bf 23}
  167

\bibitem{Dominguez1993}
Dominguez R and Staebler G 1993 {\em Nuclear Fusion\/} {\bf 33} 51

\bibitem{Paccagnella1990}
Paccagnella R, Romanelli F and Briguglio S 1990 {\em Nuclear Fusion\/} {\bf 30}
  545

\bibitem{Frojdh1992}
Frojdh M, Liljestrom M and Nordman H 1992 {\em Nuclear Fusion\/} {\bf 32} 419

\bibitem{Dong1994}
Dong J~Q, Horton W and Dorland W 1994 {\em Physics of Plasmas\/} {\bf 1} 3635

\bibitem{Dong1995}
Dong J~Q and Horton W 1995 {\em Physics of Plasmas\/} {\bf 2} 3412

\bibitem{Du2014}
Du H, Wang Z~X, Dong J~Q and Liu S~F 2014 {\em Physics of Plasmas\/} {\bf 21}
  052101

\bibitem{Du2015}
Du H, Wang Z~X and Dong J~Q 2015 {\em Physics of Plasmas\/} {\bf 22} 022506

\bibitem{Fu1997}
Fu X~Y, Dong J~Q, Horton W, Ying C~T and Liu G~J 1997 {\em Physics of
  Plasmas\/} {\bf 4} 588

\bibitem{Catto1978}
Catto P~J 1978 {\em Plasma Physics\/} {\bf 20} 719

\bibitem{Frieman1982}
Frieman E~A and Chen L 1982 {\em The Physics of Fluids\/} {\bf 25} 502

\bibitem{Beer1995}
Beer M~A, Cowley S~C and Hammett G~W 1995 {\em Physics of Plasmas\/} {\bf 2}
  2687

\bibitem{Barnes_jcp_2019}
Barnes M, Parra F and Landreman M 2019 {\em Journal of Computational Physics\/}
  {\bf 391} 365

\bibitem{GonzalezJerez_jpp_2022}
Gonz{\'{a}}lez-Jerez A, Xanthopoulos P, Garc{\'{\i}}a-Rega{\~{n}}a J~M, Calvo
  I, Alcus{\'{o}}n J, Navarro A~B, Barnes M, Parra F~I and Geiger J 2022 {\em
  Journal of Plasma Physics\/} {\bf 88} 905880310

\bibitem{Plunk2014}
Plunk G~G, Helander P, Xanthopoulos P and Connor J~W 2014 {\em Physics of
  Plasmas\/} {\bf 21} 032112

\bibitem{Zocco2016}
Zocco A, Plunk G~G, Xanthopoulos P and Helander P 2016 {\em Physics of
  Plasmas\/} {\bf 23} 082516

\bibitem{Zocco2018}
Zocco A, Xanthopoulos P, Doerk H, Connor J~W and Helander P 2018 {\em Journal
  of Plasma Physics\/} {\bf 84} 715840101

\bibitem{Rodriguez2025}
Rodr\'{\i}guez E and Zocco A 2025 {\em Journal of Plasma Physics\/} {\bf 91}
  E21

\bibitem{Parisi2020}
Parisi J~F, Parra F~I, Roach C~M, Giroud C, Dorland W, Hatch D~R, Barnes M,
  Hillesheim J~C, Aiba N, Ball J, Ivanov P~G and {JET contributors} 2020 {\em
  Nuclear Fusion\/} {\bf 60} 126045

\bibitem{Biglari1989}
Biglari H, Diamond P~H and Rosenbluth M~N 1989 {\em Physics of Fluids B: Plasma
  Physics\/} {\bf 1} 109

\bibitem{Landau1946}
Landau L 1946 {\em J. Phys. USSR\/} {\bf 10} 25

\bibitem{Fried_book}
Fried B~D and Conte S~D 1961 {\em The Plasma Dispersion Function\/} (New York:
  Academic Press)

\end{thebibliography}

\providecommand{\noopsort}[1]{}\providecommand{\singleletter}[1]{#1}%
\providecommand{\newblock}{}

\end{document}